\documentclass[superscriptaddress, aps, prl, twocolumn, showpacs, nofootinbib, longbibliography, showkeys, 10pt]{revtex4-2}
\usepackage{amsmath,amssymb,amsthm}
\usepackage{easybmat}
\usepackage[colorlinks=true,citecolor=blue,urlcolor=blue, linkcolor = magenta]{hyperref}
\usepackage[pdftex]{graphicx}
\usepackage{times,txfonts}
\usepackage{braket}
\usepackage{color}
\usepackage{natbib}
\newcommand{\be}{\begin{equation}}
	\newcommand{\ee}{\end{equation}}
\newcommand{\ba}{\begin{eqnarray}}
	\newcommand{\ea}{\end{eqnarray}}

\begin{document}
	
	%\preprint{APS/123-QED}
	\title{Quantum Thermal Analogs of Electric Circuits: A Universal Approach}

\author{Devvrat Tiwari\textsuperscript{}}
\email{devvrat.1@iitj.ac.in}%Lines break automatically or can be forced with \\
    \affiliation{Indian Institute of Technology Jodhpur-342030, India\textsuperscript{}}
       \author{Samyadeb Bhattacharya} 
         \email{samyadeb.b@iiit.ac.in}
        \affiliation{Center for Quantum Science and Technology and Center for Security Theory and Algorithmic Research, International Institute of Information Technology, Gachibowli, Hyderabad-500032, India\textsuperscript{}}
	\author{Subhashish Banerjee\textsuperscript{}}
	\email{subhashish@iitj.ac.in }
	 \affiliation{Indian Institute of Technology Jodhpur-342030, India\textsuperscript{}}


\date{\today}% It is always \today, today,

\begin{abstract}
	In this work, we develop a panoramic schematic of quantum thermal analogs of electric circuits in the steady state regime. We establish the foundations of said premise by defining the analogs of Kirchhoff's laws for heat currents and temperature gradients, as well as a quantum thermal step transformer. Using this, we develop two novel quantum thermal circuits, \textit{viz.}, quantum thermal super Wheatstone bridge and quantum thermal adder circuit, paving the way for the corresponding integrated circuits. We further show that our approach encompasses various circuits like thermal diode, transistor, and Wheatstone bridge. This sheds new light on the present architecture of quantum device engineering. 
\end{abstract}

%Use showkeys class option if keyword display is desired
\maketitle

%\tableofcontents

\textit{Introduction---}All quantum systems are inherently subject to interactions with their surrounding environment, which the theory of open quantum systems addresses~\cite{weiss, breuer, banerjee2018open}. Recent technological advancements have elevated the importance of these interactions, particularly in quantum thermodynamics, where understanding non-equilibrium processes and energy flow at the quantum level has become essential for developing quantum technologies~\cite{Gemmer2009, Binder_book, Kosloff2013, Tiwari_strong, Tiwari_hmf, Tiwari2023}.

Logical computational circuits are crucial for the development of quantum computers \citep{gates,gates1,gates2}, making quantum circuitry foundational in quantum device engineering. Drawing from classical circuit design, identifying quantum equivalents of diodes, transistors, resistors, and inductors is vital for creating a quantum microprocessor. A promising approach involves using heat currents to create quantum versions of electrical and thermal devices~\cite{circuit1,circuit2,circuit3,circuit4,circuit5,circuit6,circuit7,circuit8,circuit9,togan, maity2024, circuit10,circuit11,circuit12,circuit13,circuit14,circuit15,circuit16,circuit17,circuit18,circuit19,circuit20,circuit21,circuit22,circuit23,circuit24,circuit25,circuit26,circuit27,circuit28, XIE2023106575, hegde2024, QWB, thermal_diode, NM_Myers_thermal_devices}. Analyzing heat flow patterns in quantum networks is essential to model these quantum thermoelectric devices, focusing on identifying heat current behavior~\cite{ohmic, Tanimura_2016}. From the advent of the formal theory of open quantum systems, an important achievement of which was the establishment of the Gorini-Kossakowski-Sudarshan-Lindblad (GKSL) master equation~\citep{lindblad, GKS}, modeling quantum thermal devices has been one of its most promising applications \citep{alicki,lendi}, where heat currents are manipulated to efficient effect. This paper offers a novel quantum thermoelectric network theory, providing the fundamental laws on heat currents and temperature gradients, playing the equivalent roles of electric current and voltage differences, respectively. 

We consider the thermoelectric quantum networks from the backdrop of Markovian dynamics~\cite{breuer,lindblad, banerjee2018open}. The model involves multiple interacting qubits influenced by non-interacting thermal environmental modes, with the baths considered as collections of harmonic oscillators~\cite{breuer}. In this backdrop, the resulting reduced dynamics of the system demonstrates a one-way information flow from the system to the environment, monotonically leading the qubit toward its corresponding thermal equilibrium. It is also important to mention that all our investigations are done in the steady state region, which naturally occurs for such Markovian dynamics. This, in turn, ensures the stability of the circuit models.   

We model a quantum thermal resistor identified by the strength of the interaction Hamiltonian to find the heat current between two or more qubits. We establish the laws of a quantum thermal analog of an electric circuit, analogous to Kirchhoff's current and voltage laws, treating a qubit as the node. We also prove the existence of a quantum thermal step transformer. We develop two novel quantum thermal circuits---a quantum thermal super Wheatstone bridge, and a thermal adder circuit (motivated by the operational amplifier circuit, where the voltage output is the algebraic sum of all the inputs)---and use the above framework to study them.
It is shown that the thermal diode, transistor, and the standard Wheatstone bridge circuits fall into the framework constructed here, the details of which are presented in Supplemental Material~\cite{supplemental}. The framework introduced is the first of its kind, providing a fundamental tool to build a comprehensive quantum thermal circuit theory. The fact that many thermal versions of important classical electric circuits can be studied in a unified manner under the umbrella of this framework will interest a broader audience in the juxtaposition of quantum thermodynamics and electric circuits.

\begin{figure*}
    \centering
    \includegraphics[width=1\linewidth]{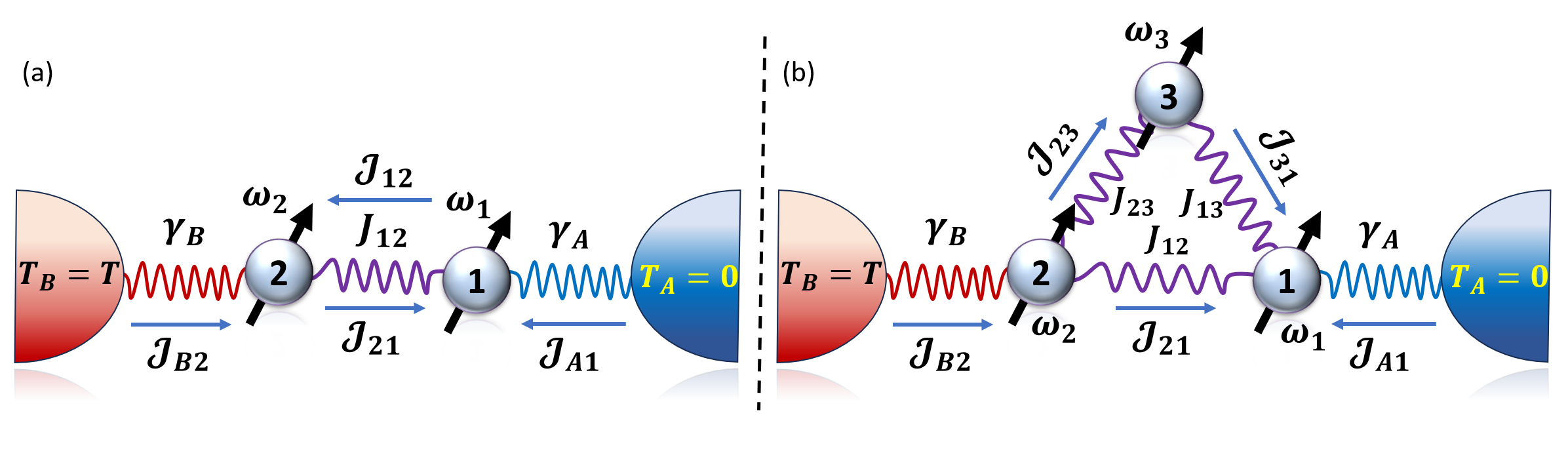}
    \caption{Schematic diagrams of the (a) quantum thermal resistor and (b) three-qubit quantum thermal circuit. The quantum thermal current from source $j$ to qubit $k$, taking qubit $k$ as a node, is $\mathcal{J}_{jk}$.}
    \label{fig_thermal_circuits}
\end{figure*}

\textit{Quantum thermal resistor and current law---}Our program begins by constructing a model for a quantum thermal resistor. By analyzing the flow of heat current through a quantum system between two thermal baths with differing temperatures, we seek to establish a relationship akin to Kirchhoff's laws. Our goal is to understand if, and under what conditions, temperature gradients and heat currents in quantum thermal devices can relate in a manner analogous to these fundamental circuit principles. The quantum thermal resistor is made up of two qubits, Fig.~\ref{fig_thermal_circuits}(a). 
Qubits 1 and 2 are under the influence of bosonic baths at temperatures zero and $T$, respectively. The Hamiltonian of this system (for $\hbar = k_B = 1$) is
$H_S = \sum_{j=1}^2\frac{\omega_j}{2}\sigma_j^z + J_{12}\left(\sigma^x_1\sigma^x_2 + \sigma^y_1\sigma^y_2\right),$
where $\sigma^i~(i = x, y, z)$ are the Pauli matrices. $J_{12}$ is the interaction strength between the two qubits and models the quantum thermal resistor. 
The dynamics of the system is governed by 
%the GKSL master equation of the form
\begin{align}
    \frac{d\rho}{dt} &= -i[H_S, \rho] + \mathcal{D}_{A1}\left(\rho\right) + \mathcal{D}_{B2}\left(\rho\right) = \mathcal{L}\left(\rho\right) ,
    \label{qtr_master_eq}
\end{align}
where $\mathcal{D}_{jk}(\rho) = \gamma_j(\tilde N_{jk} + 1)\left(\sigma^-_k\rho\sigma^+_k - \frac{1}{2}\left\{\sigma^+_k\sigma^-_k, \rho\right\}\right) + \gamma_j \tilde N_{jk}\left(\sigma^+_k\rho\sigma^-_k - \frac{1}{2}\left\{\sigma^-_k\sigma^+_k, \rho\right\}\right)$, with $\sigma^{\pm}_k = \frac{1}{2}\left(\sigma^x_k \pm i\sigma^y_k\right)$, and $\tilde N_{jk} = \frac{1}{e^{\beta_j\omega_k} - 1}$, with $\beta_j = T_j^{-1}$ (here, $j = A, B$ and $k = 1, 2$). $\gamma_A$ and $\gamma_B$ are the dissipative factors. $\mathcal{L}$ denotes the right-hand side of the master equation. We assume that each of the qubits is locally interacting with its associated bath. The steady state of the system is given by the condition $\frac{d\rho^{SS}}{dt} = 0$, and is
\begin{align}
    \rho^{SS} = \sum_{i=0}^3\frac{\alpha_{ii}}{\eta_{ii}}\ket{i}\bra{i} + \frac{\alpha_{12}\ket{1}\bra{2} + \alpha^*_{12}\ket{2}\bra{1}}{\eta_{12}} + \frac{e^{\frac{\omega_2}T}}{\left(1 + e^{\frac{\omega_2}T}\right)}\ket{3}\bra{3},
    \label{qtr_steady_state}
\end{align}
where $\alpha_{ij}$ and $\eta_{ij}$ are given in~\cite{supplemental} and $\ket{i}$ is such that $\ket{0} = \ket{0}_1\ket{0}_2, \ket{1} = \ket{0}_1\ket{1}_2, \ket{2} = \ket{1}_1\ket{0}_2$, and $\ket{3} = \ket{1}_1\ket{1}_2$  ($\ket{\cdot}_k$ denotes qubit $k$). The quantum heat currents, in the steady state condition, from the bath $j$ to the qubit $k$ ($\mathcal{J}_{jk}$) and from qubit $l$ to qubit $k$ ($\mathcal{J}_{lk}$) are given by 
\begin{align}
    \mathcal{J}_{jk} = {\rm Tr} \left[H_k\mathcal{D}_{jk}\left(\rho^{SS}\right)\right]~~\text{and}~~\mathcal{J}_{lk} = i{\rm Tr}\left\{\rho^{SS}[H_{lk}, H_k]\right\},
    \label{heat_current_bath_system}
\end{align}
respectively~\cite{supplemental}. The above heat currents are derived from the time derivative of heat transferred to the qubit, providing heat current between individual components of the circuit. Considering a spin-chain, the above heat current formula becomes equivalent to the total heat current ${\rm Tr}\left[H_S \mathcal{D}_{jk}(\rho^{SS})\right]$ for equal qubit frequencies $\omega_j$ (see~\cite{supplemental}), which is fundamental to studies in thermal conductivity~\cite{lendi, Dhar_Saito_Hanggi}. Further, the heat current between the nodes is also related to the spin current~\cite{Poletti_2018, Poulsne_2022, Poulsen_2024, Mendoza_2024}, as shown in~\cite{supplemental}.

Using the state $\rho^{SS}$ from Eq. (\ref{qtr_steady_state}), the current $\mathcal{J}_{A1}$ from bath $A$ to qubit 1 is  
    $\mathcal{J}_{A1} = {\rm Tr}\left[H_1\mathcal{D}_{A1}\left(\rho^{SS}\right)\right] = - \gamma_A\omega_1 \left(\frac{\alpha_{00}}{\eta_{00}} + \frac{\alpha_{11}}{\eta_{11}}\right).$
Pertinently, since the parameters $\gamma_A\ge0$, $\alpha_{00}/\eta_{00}\ge 0$ and $\alpha_{11}/\eta_{11}\ge 0$, the current $\mathcal{J}_{A1}$ negative. The rationale behind $\mathcal{J}_{A1}$ being negative is that qubit 1 is surrounded by a zero-temperature bath, and negative $\mathcal{J}_{A1}$ indicates that the heat flow is directed from qubit 1 toward the thermal reservoir. Therefore, reservoir $A$ acts as a ``ground" to which the quantum thermal resistor is connected. Furthermore, the current $\mathcal{J}_{21}$ from qubit 2 to 1 is given by 
$\mathcal{J}_{21} = \frac{2i J_{12}\omega_1\left(\alpha_{12} - \alpha_{12}^*\right)}{\eta_{12}} = \frac{-4 J_{12} \omega_1 \Im \left(\alpha_{12}\right)}{\eta_{12}},$
where $\Im(*)$ denotes imaginary part. Interestingly, we find 
$\mathcal{J}_{A1} = -\left(\frac{\gamma_A\omega_1x_0}{\gamma_B}\right)\frac{\alpha_{00}}{\eta_{00}} = -\mathcal{J}_{21}$, resulting in
$
    \mathcal{J}_{A1} + \mathcal{J}_{21} = 0.
$
This is crucial as it allows qubit 1 to be considered as a junction node, leading to the visualization of the above equation as the quantum thermal version of \textit{Kirchhoff's current law}. In fact, considering heat currents $\mathcal{J}_{jk}$'s from $n$ different sources to a qubit $k$, it can be proved that in the steady state condition (see~\cite{supplemental}), the sum of all the heat currents is zero, i.e.,
\begin{align}
    \sum_{j = 1}^n \mathcal{J}_{jk} = 0.
    \label{kirchhoff_current_law}
\end{align}%
Similarly, we find that the currents with qubit 2 as the junction node are given by $\mathcal{J}_{B2} = {\rm Tr}\left[H_2\mathcal{D}_{B2}\left(\rho^{SS}\right)\right] = \left(\frac{\gamma_A\omega_2x_0}{\gamma_B}\right)\frac{\alpha_{00}}{\eta_{00}}$,
and 
$\mathcal{J}_{12} = -\left(\frac{\gamma_A\omega_2x_0}{\gamma_B}\right)\frac{\alpha_{00}}{\eta_{00}}.$
It can be easily verified that $\mathcal{J}_{B2} + \mathcal{J}_{12} = 0$, which shows the characteristics of the quantum thermal version of Kirchhoff's current law. 

\textit{Quantum thermal transformer.---} Interestingly, the heat currents $\mathcal{J}_{21}$ and $\mathcal{J}_{12}$ are interconnected by the relation $
\frac{\mathcal{J}_{21}}{\omega_1} =  - \frac{\mathcal{J}_{12}}{\omega_2},$ or $\left|\frac{\mathcal{J}_{21}}{\mathcal{J}_{12}}\right| = \frac{\omega_1}{\omega_2} = \left|\frac{\mathcal{J}_{A1}}{\mathcal{J}_{B2}}\right|. $
In general, for an interaction $H_{jk} = J_{jk}\left(\sigma^x_j\sigma^x_k + \sigma^y_j\sigma^y_k\right)$ between two qubits $j$ and $k$, the current between them can be shown to be~\cite{supplemental}
\begin{align}
    \frac{\mathcal{J}_{jk}}{\omega_k} = -\frac{\mathcal{J}_{kj}}{\omega_j}.
    \label{eq_thermal_transformer}
\end{align}
The above relation mimics the quantum thermal version of a step transformer, where the ratio of currents between two coils depends on the number of turns in each coil. Here, the ratio of quantum heat currents depends on the transition frequencies $\omega_j$ and $\omega_k$ of the qubits $j$ and $k$, respectively, thereby allowing their manipulation. 

\textit{Quantum thermal voltage---}We now develop the quantum analog of the thermal voltage driving the thermal current. To this end, we determine the effective temperature of qubits 1 and 2 in the steady state. The reduced state of qubit 1 from the steady state, Eq.~\ref{qtr_steady_state}, is given by
$\rho^{SS}_1 = \left(\rho^{SS}_{00} + \rho^{SS}_{11}\right) \ket{0}\bra{0} + \left(\rho^{SS}_{22} + \rho^{SS}_{33}\right)\ket{1}\bra{1}.$
The effective temperature $T_1$ of the qubit 1 is found by comparing the state $\rho^{SS}_1$ with the state $e^{-H_1/T_1}/{\rm Tr}\left(e^{-H_1/T_1}\right)$, where $H_1 = \frac{\omega_1}{2}\sigma^z_1$, leading to 
$T_1 = \omega_1/\log\left(\frac{1}{\rho^{SS}_{00} + \rho^{SS}_{11}} - 1\right) = \omega_1/\log\left(\frac{\gamma_B\eta_{00}}{x_0\alpha_{00}} - 1\right).$
Note that $T_1$ can be envisaged as the thermal potential difference between qubit 1 and bath $A$, that is, $V_{1A} = T_1 - 0 = T_1$. This potential difference drives the thermal current $\mathcal{J}_{A1}$, which explicitly in terms of $T_1$ can be written as 
$\mathcal{J}_{A1} = \frac{-\gamma_A\omega_1}{1 + e^{\omega_1/T_1}}.$
In the low and high-temperature limits, this is 
\begin{align}
    \mathcal{J}_{A1} = \begin{cases}
        \frac{-\gamma_A\omega_1}{2} & \text{for}~~  T_1\gg 0,\\
        -\gamma_A\omega_1e^{-\omega_1/T_1} & \text{for}~~ T_1 \sim 0.
    \end{cases}
    \label{eq_relation_between_current_and_voltage}
\end{align}
\begin{figure}
    \centering
    \includegraphics[width = 1\columnwidth]{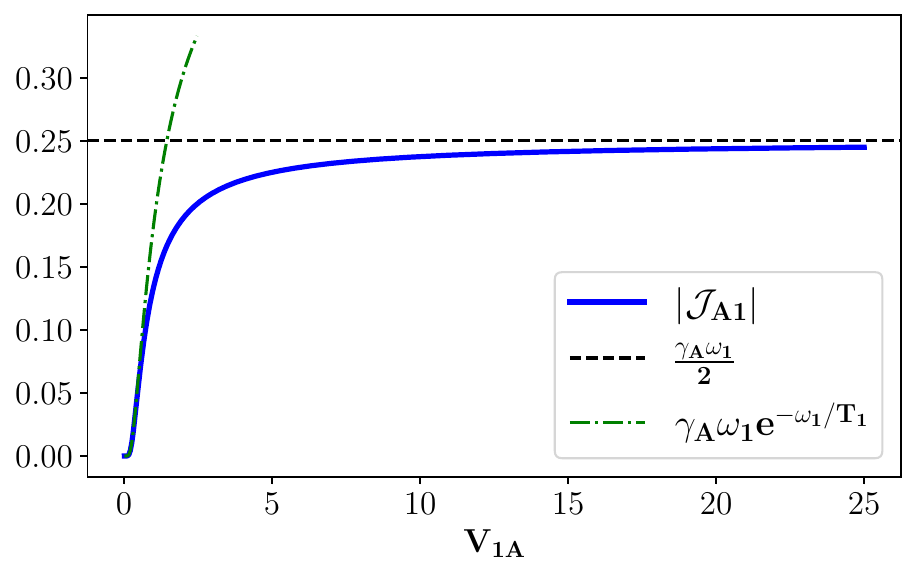}
    \caption{Variation of the thermal current $|\mathcal{J}_{A1}|$ between the qubit 1 and the bath $A$ as a function of the thermal potential difference $V_{1A} = T_1$ for $\gamma_A = 0.5$ and $\omega_1 = 1.0$.}
    \label{fig_Q1vsT1}
\end{figure}%
This has been depicted in Fig.~\ref{fig_Q1vsT1}. Similarly, one can find the reduced state of qubit 2 and its corresponding effective temperature, which is 
    $T_2 = \omega_2/\log\left(\frac{1}{\rho^{SS}_{00} + \rho^{SS}_{22}} - 1\right).$
The potential difference between the bath $B$ and the qubit 2 can now be written as $V_{B2} = T - T_2$. Further,
$\mathcal{J}_{B2} = \frac{\gamma_B\omega_2}{2}\left[-1 + \coth\left(\frac{\omega_2}{2T}\right)\tanh\left(\frac{\omega_2}{2\left\{T-V_{B2}\right\}}\right)\right].$
It can be pointed out here that the thermal current $\mathcal{J}_{B2}$ does not explicitly depend only on the thermal potential difference $V_{B2}$ but also on the temperature ($T$) of the bath $B$. 
\begin{figure}
    \centering
    \includegraphics[width = 1\columnwidth]{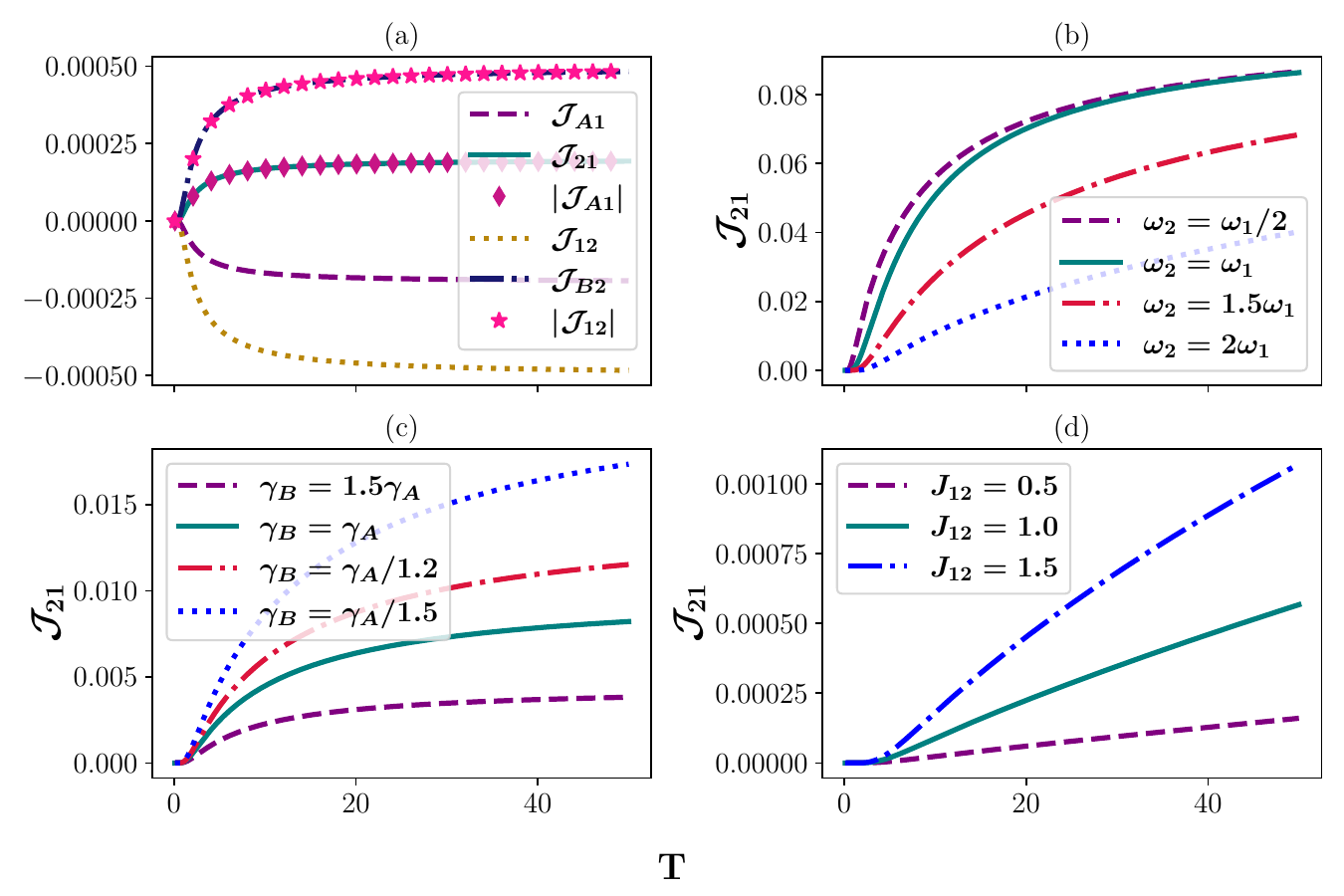}
    \caption{Variation of heat currents $\mathcal{J}_{jk}$ in (a), and heat current $\mathcal{J}_{21}$ from qubit 2 to 1 in (b), (c), and (d) with temperature $T$ of bath $B$ for the two-qubit quantum thermal resistor model. (a) $\omega_1 = 2, \omega_2 = 2.5, \gamma_A = 0.01, \gamma_B = 0.05$, and $J_{12} = 1.0$; (b) $\omega_1 = 5, \gamma_A = 0.01, \gamma_B = 0.005$, and $J_{12} = 1.0$; (c) $\omega_1 = 2, \omega_2 = 5, \gamma_A = 0.01$, and $J_{12} = 1.0$; (d) $\omega_1 = 1.0, \omega_2 = 15$, and $\gamma_A = \gamma_B = 0.01$. }
    \label{fig_Q_vs_T}
\end{figure}
The heat currents $\mathcal{J}_{jk}$ discussed here are plotted in Fig.~\ref{fig_Q_vs_T} as a function of temperature $T$. The heat current can be seen to increase linearly in certain parameter regimes. Thus, for example, in Fig.~\ref{fig_Q_vs_T}(d), for $\omega_2\gg\omega_1$, we find the current $\mathcal{J}_{21}$ increases almost linearly with $T$ (particularly, for $J_{12} = 0.5$ and $J_{12} = 1$), satisfying the condition for Ohm's law. However, in general, the heat current exhibits nonlinear behavior. Further, Kirchhoff's current law is satisfied universally.

\textit{Law of thermal potentials---}We now move on to study quantum thermal circuits made up of multiple qubits. A schematic diagram is given in Fig. ~\ref{fig_thermal_circuits}(b), with the Hamiltonian $\widetilde H_S = \frac{1}{2}\sum_{i = 1}^3\omega_i\sigma^z_i + \sum_{l, k = 1, l<k}^3J_{lk}\left(\sigma^x_l\sigma^x_k + \sigma^y_l \sigma^y_k\right)$, where qubits $l$ and $k$ interact with each other with the interaction strength $J_{lk}$ and $\omega_k$ being the transition frequency of the $k$-th qubit. Qubits 1 and 2 are under the influence of the bath. We compute the steady-state $\tilde \rho^{SS}$ of this system using Eq.~\eqref{qtr_master_eq} by replacing $H_S$ with $\widetilde H_S$.    
\begin{figure}
    \centering
    \includegraphics[width = 1\linewidth, height = 5cm]{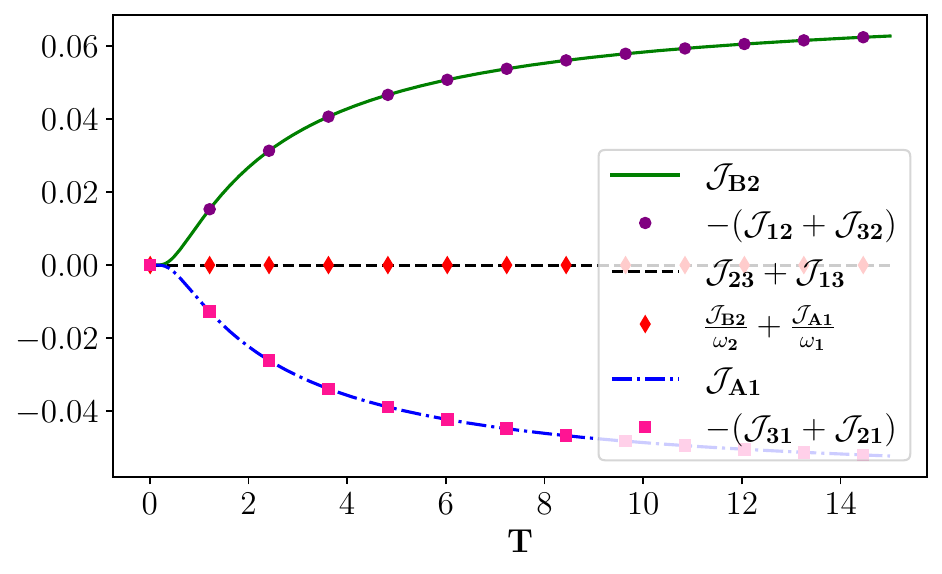}
    \caption{Variation of quantum heat currents in the case of the three-qubit quantum thermal circuit in steady state. The parameters are: $\omega_1 = 1.25, \omega_2 = 1.5, \omega_3 = 1.75, J_{12} = 1.0, J_{13} = 0.5, J_{23} = 0.75, \gamma_A = 0.1, \gamma_B = 0.05$.}
    \label{fig_QvsT_three_qubit}
\end{figure}
The variation of the quantum heat currents in this circuit is shown in Fig.~\ref{fig_QvsT_three_qubit}. It can be verified from this figure that Kirchhoff's current law holds true at a junction node.
Further, the effective temperatures of the qubits 1, 2, and 3 are given by
$ T_1 = \omega_1/\log\left(\frac{1}{\tilde \rho^{SS}_{00} + \tilde \rho^{SS}_{11} + \tilde \rho^{SS}_{22} + \tilde \rho^{SS}_{33}} - 1\right)$, 
$T_2 = \omega_2/\log\left(\frac{1}{\tilde \rho^{SS}_{00} + \tilde \rho^{SS}_{11} + \tilde \rho^{SS}_{44} + \tilde \rho^{SS}_{55}} - 1\right)$, and
$T_3 = \omega_3/\log\left(\frac{1}{\tilde \rho^{SS}_{00} + \tilde \rho^{SS}_{22} + \tilde \rho^{SS}_{44} + \tilde \rho^{SS}_{66}} - 1\right)$, respectively.
The corresponding thermal potential differences are
$V_{1A} = T_1 - 0;  V_{21} = T_2 - T_1;  V_{32} = T_3 - T_2; V_{13} = T_1 - T_3;  V_{B2} = T - T_2.$
\begin{figure*}
    \centering
    \includegraphics[width=0.85\linewidth]{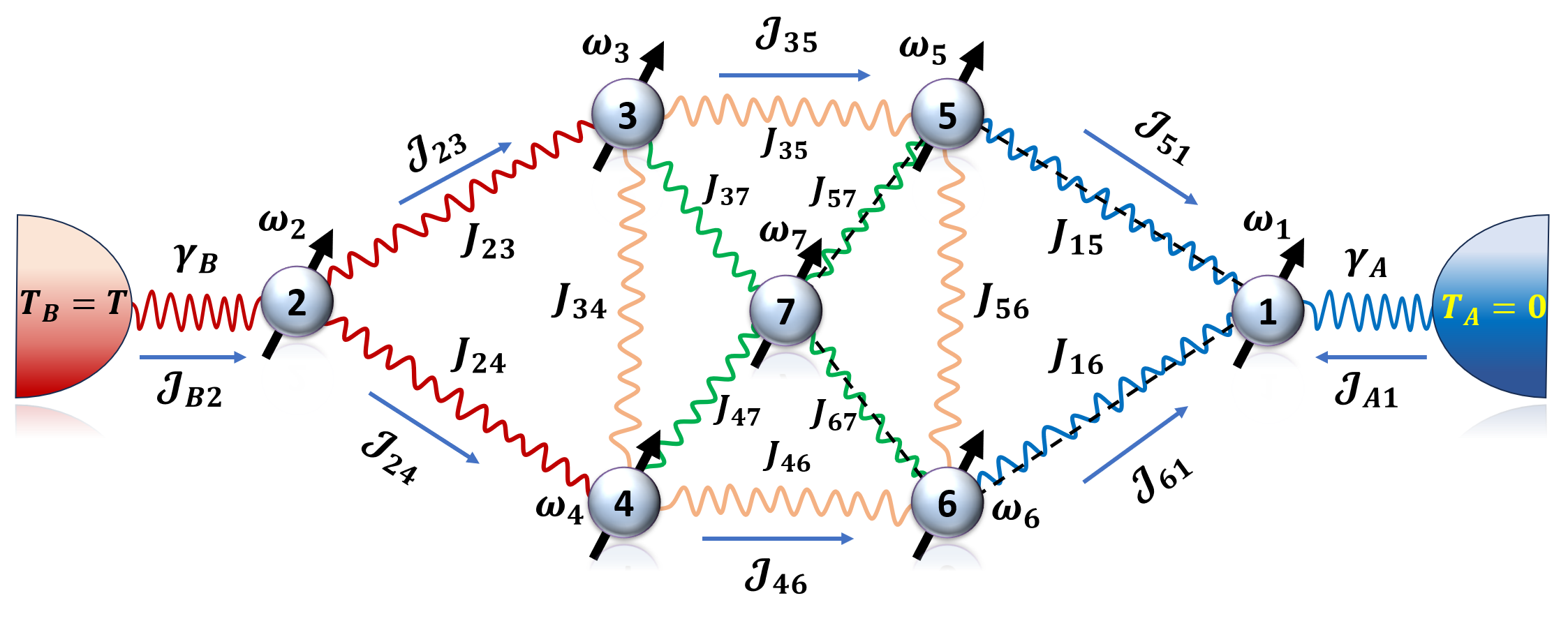}
    \caption{A schematic diagram of the quantum thermal super Wheatstone Bridge.}
    \label{fig_quantum_thermal_super_Wheatstone_bridge}
\end{figure*}%
Using this, we get a relation $V_{21} + V_{32} + V_{13} = 0$ inside the loop of the three qubits, resembling \textit{Kirchhoff's second law of an electric circuit}, stating that the sum of potential differences around a closed loop (a closed sequence of interactions connecting qubits, returning to the initial qubit) in an electric circuit is zero. The relation between the thermal currents $\mathcal{J}_{A1}, \mathcal{J}_{B2}$ and the potential differences $V_{1A}, V_{B2}$ is the same as in the case of the previous circuit.
The currents from qubit 3 to 1 and from qubit 2 to 1 are given by  $\mathcal{J}_{31} = -4J_{13}\omega_1\left[\Im{(\tilde \rho^{SS}_{14})} + \Im{(\tilde \rho^{SS}_{36})}\right]$ and $\mathcal{J}_{21} = -4J_{12}\omega_{1}\left[\Im{(\tilde \rho^{SS}_{24})} + \Im{(\tilde \rho^{SS}_{35})}\right]$, respectively. This leads to the relation~\cite{supplemental}
\begin{align}
    \mathcal{J}_{31} = \frac{4J_{13}}{\tilde f_{13}}V_{13} = -\frac{\omega_1\mathcal{J}_{13}}{\omega_3},
    \label{eq_three_qubit_q13_vs_v13}
\end{align}
where the $\tilde f_{13}$ is a function of $\omega_j, T_j, J_{ij}, T$, and $\gamma_k$ (for $i = 1, 2; j = 1, 2, 3$ and $k = A, B$). Similarly, one can find out the various functions, such as $\tilde f_{12}$, and $\tilde f_{23}$, which can be used to identify the corresponding relation of the quantum thermal currents $\mathcal{J}_{12}, \mathcal{J}_{21}, \mathcal{J}_{23}$, and $\mathcal{J}_{32}$ with the quantum thermal potentials $V_{12}$ and $V_{23}$. 

We use the above framework to develop novel circuits, to which we coin the names quantum thermal super Wheatstone bridge and quantum thermal adder. Further, thermal devices discussed in recent times, for example, a Wheatstone bridge, a diode, and a quantum thermal transistor, are presented using our framework in~\cite{supplemental}.  

\begin{figure}
    \centering
    \includegraphics[width=1\linewidth]{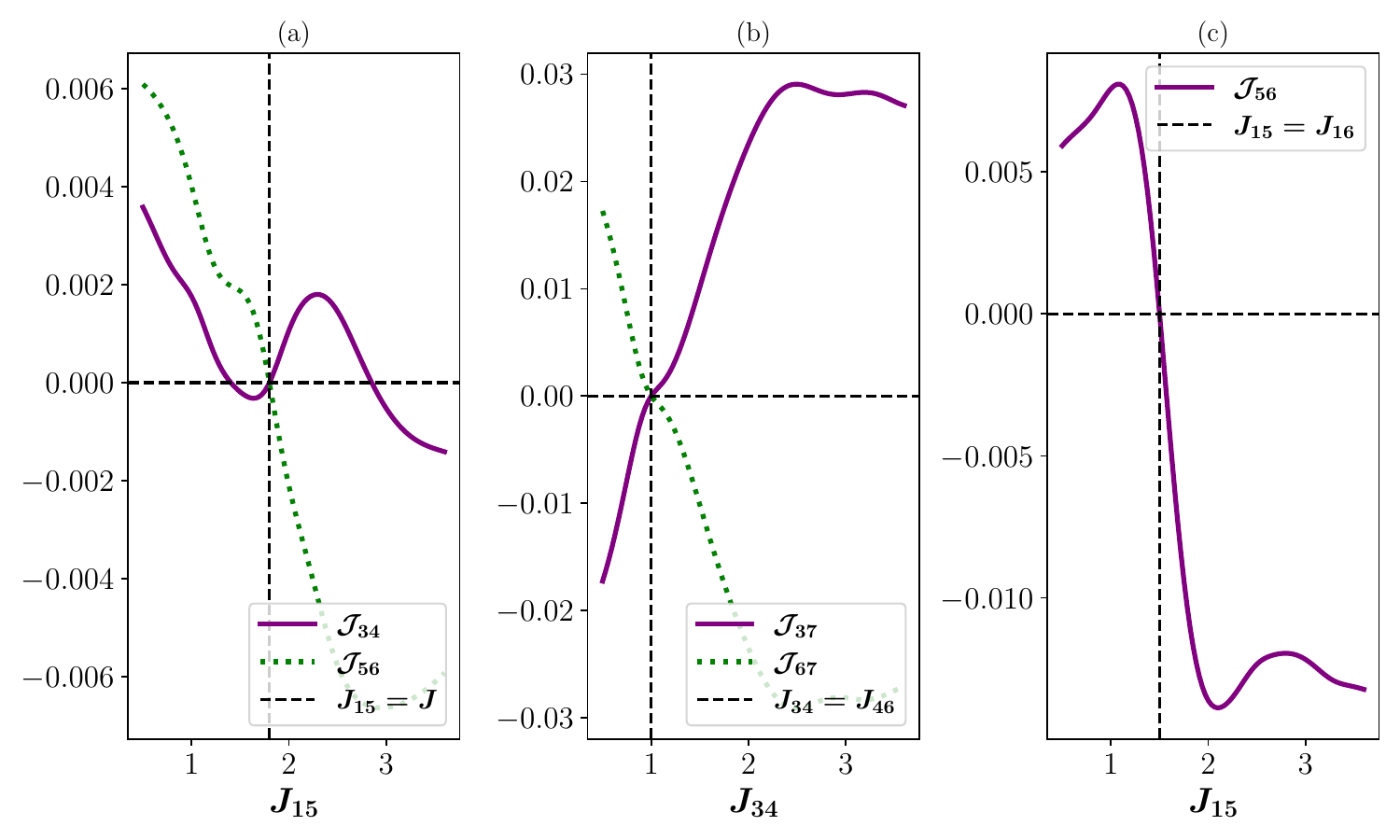}
    \caption{Variation of thermal currents $\mathcal{J}_{jk}$ with interaction strengths $J_{jk}$. (a), (b) and (c) depict conditions 1, 2, and 3 of the super Wheatstone bridge, mentioned in the text below. In all the plots, unless conditioned for the balanced state of the bridge, the values of $J_{jk}$'s are of the order of $\omega_j$'s, which are randomly taken between 1 and 2. Further, $T= 10, \gamma_A = 0.1$, and $\gamma_B = 0.05$.}
    \label{fig_super_Wheatstone_bridge}
\end{figure}

\textit{Quantum thermal super Wheatstone bridge---}We discuss the circuit for a quantum thermal super Wheatstone Bridge, Fig.~\ref{fig_quantum_thermal_super_Wheatstone_bridge}, with Hamiltonian 
$
 H_S^{SW} =  \sum_{j = 1}^7 \frac{\omega_j}2 \sigma^z_j + H_{15} + H_{16} + H_{23} + H_{24} + H_{34} + H_{35} + H_{37} + H_{46} + H_{47} + H_{56} + H_{57} + H_{67}
$, where $H_{jk} = J_{jk}\left(\sigma^x_j\sigma^x_k + \sigma^y_j\sigma^y_k\right)$. The dynamics of the system is governed by Eq.~\eqref{qtr_master_eq}, by replacing $H_S$ with $H_S^{SW}$. The geometry of this circuit makes it very interesting. We discuss three important facets of this circuit here. In condition 1, the circuit is made symmetrical, such that all the $\omega_j$'s as well as all the $J_{jk}$'s are the same ($J$). In this condition, we observe that the currents $\mathcal{J}_{34}$ and $\mathcal{J}_{56}$ become zero, see Fig.~\ref{fig_super_Wheatstone_bridge}(a).
In condition 2, we explore the scenario where the center qubit 7 is unaffected by the thermal transport around it. To do so, we allow the input and output of heat from single connections to the square made by qubits 3, 4, 5, and 6, thereby making $J_{23} = J_{61} = 0$. Further, we allow only two connections to qubit $7$, for example, with qubit 3 and 6, implying $J_{57} = J_{47} = 0$. The balance condition is obtained when $J_{34} = J_{46}$ and $J_{35} = J_{56}$ ($J_{34}$ need not be equal to $J_{35}$), together with $\omega_3 = \omega_6 = \omega_7$. In this condition, see Fig.~\ref{fig_super_Wheatstone_bridge}(b), the currents $\mathcal{J}_{37} = \mathcal{J}_{67} = 0$, making the qubit 7 unaffected by the thermal transport around it. Interestingly, if we increase the number of qubits between qubits 3 and 6, keeping their $\omega_i$'s equal, they will also remain unaffected by the thermal transport around them~\cite{supplemental}. Condition 3 of the super Wheatstone bridge is identical to the Wheatstone bridge condition~\cite{supplemental}. We can form many Wheatstone bridges inside this super Wheatstone bridge; for example, see the black dashed line between qubits 1, 5, 7, and 6. Here, we take $J_{35} = J_{37} = J_{46} = 0$ to form a particular Wheatstone bridge. The balance condition is achieved when $J_{15} = J_{16}$ and $J_{57} = J_{67}$ ($J_{15}$ need not be equal to $J_{57}$); and then the current between qubits 5 and 6, $\mathcal{J}_{56}$, becomes zero, see Fig.~\ref{fig_super_Wheatstone_bridge}(c), a classic Wheatstone bridge condition. 

\begin{figure}
    \centering
    \includegraphics[width=1\linewidth]{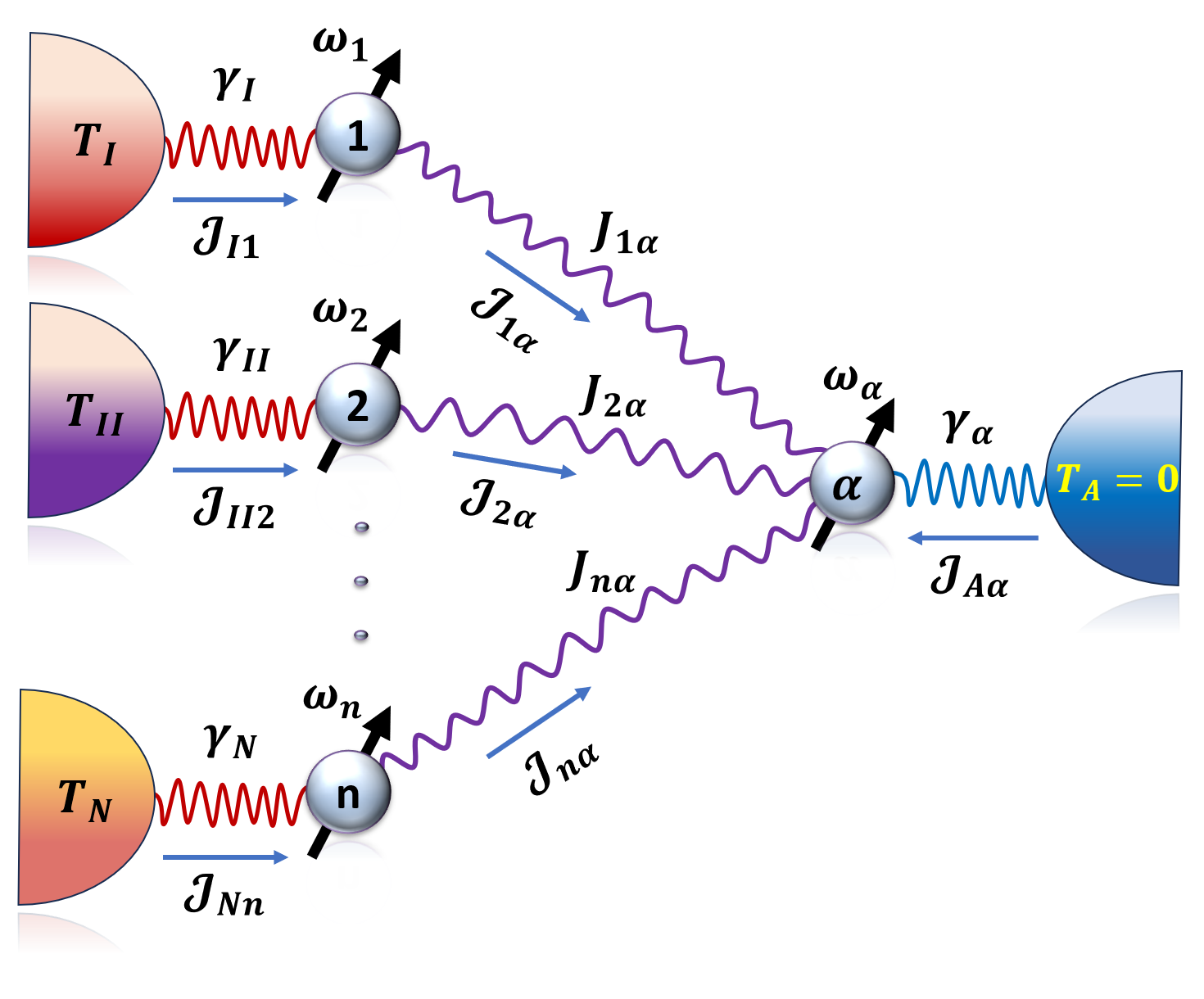}
    \caption{A schematic diagram of the quantum thermal adder circuit.}
    \label{fig_thermal_circuits_bridge_and_adder}
\end{figure}

\textit{Quantum thermal adder circuit---}We now discuss the quantum thermal adder circuit, Fig.~\ref{fig_thermal_circuits_bridge_and_adder}. 
This thermal circuit is motivated by the DC operational amplifier's adder circuit, where the output voltage of the device is the sum of all the input voltages. Here, the qubits $1, 2, ..., n$ interact with the baths $I, II, ..., N$, at temperatures $T_I, T_{II}, ..., T_N$, respectively. The effective temperatures of the qubits $1, 2, ..., n$ are $T_1, T_2, ..., T_n$, respectively. 
The thermal potential difference between qubit 1 and bath $I$ is  $V_{I1} = T_I - T_1$ and similarly for the other qubits. 
Qubits $1, 2, ..., n$ also interact with qubit $\alpha$ connected to a zero temperature bath $T_A$ and the thermal potential difference between qubit $\alpha$ and bath $A$ is $V_{\alpha A} = T_{\alpha}- T_A= T_{\alpha}$, where $T_{\alpha}$ is the effective temperature of qubit $\alpha$. The Hamiltonian of the system is $H_{add} = \frac{\omega_\alpha}{2}\sigma^z_\alpha + \sum_{i = 1}^n\left[\frac{\omega_i}{2}\sigma^z_i + J_{i\alpha}\left(\sigma^x_\alpha\sigma^x_i + \sigma^y_\alpha\sigma^y_i\right)\right]$. Using the thermal current law, $\mathcal{J}_{A\alpha} = \sum_{k = 1}^n \mathcal{J}_{k\alpha}$ and Eq.~\eqref{eq_thermal_transformer}, we can write $\mathcal{J}_{A\alpha} = -\sum_{k = 1}^n\frac{\omega_\alpha}{\omega_k}\mathcal{J}_{\alpha k}$. But $\mathcal{J}_{\alpha k} = \mathcal{J}_{jk}$ (for $(j, k) = (I, 1), (II, 2), ...(N, n)$), implying the following relation 
\begin{align}
    \mathcal{J}_{A\alpha} = -\omega_\alpha\left(\frac{\mathcal{J}_{I1}}{\omega_1} + \frac{\mathcal{J}_{II2}}{\omega_2}+...+\frac{\mathcal{J}_{Nn}}{\omega_n}\right),
    \label{eq_thermal_adder_result1}
\end{align}
which for $\omega_k = \omega_\alpha \forall k$ results into $\mathcal{J}_{A\alpha} = -\left(\mathcal{J}_{I1} + \mathcal{J}_{II2}+...+\mathcal{J}_{Nn}\right)$. The above equation exactly mimics the relation between the input and output voltages of an adder circuit. The currents $\mathcal{J}_{jk}$ and $\mathcal{J}_{A\alpha}$ are related to the effective temperature of the corresponding qubits $T_k$ via
$
    \mathcal{J}_{jk} = \gamma_j\omega_k\left[-1 + \coth\left(\frac{\omega_k}{2 T_j}\right)\tanh\left(\frac{\omega_k}{2 T_k}\right)\right],
$
for $j = A, I, II, ...N$ and $k = \alpha, 1, 2, ...n$. Taking $\gamma_A = \gamma_I =...=\gamma_N$, $\omega_1=\omega_2 =...=\omega_n = \omega$, $J_{1\alpha} = J_{2\alpha}=...=J_{n\alpha}$ and $T_I = T_{II}=...=T_{N} = T$, we get a simplified relation between the effective temperatures $T_x$ for  $x = 1, 2,...,n$ and the qubit $\alpha$ ($T_\alpha$)
\begin{align}
    \tanh\left(\frac{\omega}{2T}\right)\left(N + \frac{2}{1 + e^{\omega_\alpha/T_\alpha}}\right) = N\tanh\left(\frac{\omega}{2T_x}\right).
    \label{eq_thermal_adder_result2}
\end{align} 
We observe that at lower temperatures $T$, $\frac{2}{1 + e^{\omega_\alpha/T_\alpha}} \rightarrow 0$. Interestingly, for higher values of $T$ and lower values of $\omega_\alpha$, the potential difference between bath $\mathcal{B}$ and qubits $V_{\mathcal{B}x} = T-T_x$ becomes independent of $T_\alpha$ and reduces to $V_{\mathcal{B}x} \approx \frac{T}{N+1}$. Remarkably, under certain parameter regimes, we get the condition that the sum of input thermal potentials becomes approximately equal to the output potential. For example, for two qubits 1 and 2 interacting with qubit $\alpha$, keeping $\omega_1 = \omega_2 = \omega$, $J_{1\alpha} = J_{2\alpha}$, $T_I = T_{II} = T$, and $\gamma_I = \gamma_{II} = \gamma$, we get a condition on $\omega_\alpha$ ($\omega_\alpha = 1.618N\omega$), and $\gamma_A$ ($\gamma_A = N\gamma$), such that $V_{I1} + V_{II2} = V_{\alpha A}$ with high accuracy in the high-temperature ($T$) regime. Similarly, for a higher number of qubits, $\omega_\alpha$ and $\gamma_\alpha$ can be tuned appropriately to get $V_{I1} + V_{II2} + ... + V_{Nn} = V_{\alpha A}$, in general.

\textit{Conclusion---}We have presented a comprehensive schematic for a quantum thermal circuit theory, paving a roadmap to construct quantum thermal integrated circuits, the important ingredients of which are quantum thermal diodes and transistors. We prove the existence of Kirchhoff-like voltage and current laws, with interaction strength playing the role of circuit resistance. We further demonstrate how heat currents can be manipulated to construct a quantum step transformer. Using this, we develop a quantum super Wheatstone bridge with many interesting balance conditions, also containing the standard quantum Wheatstone bridge, where unknown Hamiltonian strengths can be determined by the profiling of heat currents. This circuit can have a plethora of applications in process tomography, metrology, and quantum sensing. A model of a quantum thermal adder circuit for temperature gradients was also constructed, creating further opportunities to design thermal operational amplifiers and other integrated circuits. We also bring out the interrelation between the heat current developed here and the established forms of current used in the literature. Thus, for example, the spin current, which can be experimentally measured~\cite{spin_current_expt_1, spin_current_expt_2, lendi}, provides a handle for experimental determination of the heat current.
The framework of the circuits developed here can be adapted to experimental realization. Thus, for example, progress has been made towards the diode architecture with pentamethyl-disilane in an NMR register~\cite{mahesh}. Also, an experimental study of the thermodynamic uncertainty relation in the two-qubit NMR spin system, sodium fluorophosphate, was performed in~\cite{Mahesh_2020}, and an experimental setup of a quantum thermal transistor using a superconducting circuit was proposed in~\cite{Zinner_2020}. 
Along similar lines, experimental demonstration of other quantum thermal circuits discussed here could be envisaged.
Our analysis opens up several research avenues, such as studies in thermal conductivity, balance conditions of the super Wheatstone bridge, and other analogs of electric circuits.
It is thus sufficient to say that the present work offers the fundamentals of a comprehensive theory of quantum thermal analogs of electric circuits, paving the way for the corresponding integrated circuits.

\textit{Acknowledgments---}S. Bhattacharya acknowledges the support from the Ministry of Electronics and Information Technology, Government of India, under \textbf{Grant No.} 4(3)/2024-ITEA. S. Banerjee acknowledges useful discussions with T.S. Mahesh on the possible experimental realizations of quantum thermal devices. 

\bibliographystyle{apsrev4-2}
\bibliography{reference}

\onecolumngrid
\begin{center}
    \textbf{\large Supplemental Material for ``Quantum Thermal Analogs of Electric Circuits: A Universal Approach''}
\end{center}
\twocolumngrid
\section{Steady-state of quantum thermal resistor}
In the following, we discuss the quantum thermal resistor. The Hamiltonian of the system made up of two qubits (for $\hbar = k_B = 1$) is given by
\begin{align}
    H_S &= H_{1} + H_{2} + H_{12} \nonumber \\
    &=  \frac{\omega_1}{2}\sigma_1^z + \frac{\omega_2}{2}\sigma_2^z + J_{12}\left(\sigma^x_1\sigma^x_2 + \sigma^y_1\sigma^y_2\right),
    \label{qtr_ham}
\end{align}
where $\omega_1$ and $\omega_2$ are the transition frequencies for qubits 1 and 2, respectively, and $\sigma^k$'s (for $k=x,y,z$) are the Pauli spin matrices. $J_{12}$ is the interaction strength between qubits 1 and 2. To discuss the thermal transport between the two qubits and their respective baths, we find the steady state of the system. The dynamics of the system under the Born-Markov and the secular approximations are given by the Gorini-Kossakowski-Sudarshan-Lindblad (GKSL) master equation of the form 
\begin{align}
    \frac{d\rho}{dt} &= -i[H_S, \rho] + \mathcal{D}_{A1}\left(\rho\right) + \mathcal{D}_{B2}\left(\rho\right), \nonumber \\
    &= -i[H_S, \rho] + \gamma_A\left(\sigma^-_1\rho\sigma^+_1 - \frac{1}{2}\left\{\sigma^+_1\sigma^-_1, \rho\right\}\right) 
    \nonumber \\
    &+ \gamma_B\left(N_{th, B2} + 1\right)\left(\sigma_2^-\rho\sigma_2^+ - \frac{1}{2}\left\{\sigma^+_2\sigma^-_2, \rho\right\}\right) \nonumber \\
    &+ \gamma_B N_{th, B2}\left(\sigma^+_2\rho\sigma^-_2 - \frac{1}{2}\left\{\sigma^-_2\sigma^+_2, \rho\right\}\right),
    \label{appendix_qtr_master_eq}
\end{align}
where $\sigma^{\pm}_j = \frac{1}{2}\left(\sigma^x_j \pm i\sigma^y_j\right)$, and $N_{th,B2} = \frac{1}{e^{\omega_2/T_B} - 1}$ with $T_B = T$. $\gamma_A$ and $\gamma_B$ are the dissipative factors. The steady state of the system is given by the condition $\frac{d\rho^{SS}}{dt} = 0$. Here, baths $A$ and $B$ are at zero and $T$ temperatures, respectively. The form of the density matrix $\rho^{SS}$ that satisfies the said condition is  
\begin{align}
    \rho^{SS} = \begin{pmatrix}
        \frac{\alpha_{00}}{\eta_{00}} && 0 &&0 && 0\\
        0 && \frac{\alpha_{11}}{\eta_{11}} && \frac{\alpha_{12}}{\eta_{12}} && 0 \\
        0 && \frac{\alpha_{12^*}}{\eta_{12}} && \frac{\alpha_{22}}{\eta_{22}} && 0 \\
        0 && 0 && 0 &&1 - \frac{1}{1 + e^{\beta\omega_2}} + \frac{\alpha_{33}}{\eta_{33}}
    \end{pmatrix},
    \label{appendix_qtr_steady_state}
\end{align}
where
\begin{widetext}
    
\begin{align}
    \alpha_{00} &= 16J_{12}^2\gamma_B^2\left[\gamma_A + \gamma_B\coth\left(\frac{\beta\omega_2}{2}\right)\right], \nonumber \\
    \eta_{00} &= x_0 \left[\gamma_A\left\{16J_{12}^2 x_0 + \gamma_B\left(1 + e^{\beta\omega_2}\right)\left(\gamma_A^2 + 4(\omega_1 - \omega_2)^2\right)\right\}\right. \nonumber 
    +\left.\gamma_B\coth\left(\frac{\beta\omega_2}{2}\right)\left\{16 J_{12}^2x_0 + \gamma_A\gamma_B\left(1 + e^{\beta\omega_2}\right)\left(2\gamma_A + \gamma_B \coth\left(\frac{\beta\omega_2}{2}\right)\right)\right\}\right], \nonumber \\
    \alpha_{11} &= \left(x_0 - \gamma_B\right)\alpha_{00},~~~
    \eta_{11} = \gamma_B\eta_{00},~~~
    \alpha_{12} = -4 J_{12} \gamma_A\gamma_B\left[i\left\{\gamma_A + \gamma_B\coth\left(\frac{\beta\omega_2}{2}\right)\right\} + 2\left(\omega_1 - \omega_2\right)\right], \nonumber \\
    \eta_{12} &= \frac{\eta_{00}}{x_0}, \nonumber \\
    \alpha_{22} &= \gamma_B\left[16 J_{12}^2\left(e^{\beta\omega_2} - 1\right)\left(x_0 - \gamma_B\right) + \gamma_A\left\{x_0^2 + 4\left(e^{\beta\omega_2} - 1\right)^2\left(\omega_1 - \omega_2\right)^2\right\}\right], \nonumber \\
    \eta_{22} &= \frac{\eta_{00}\left(e^{\beta\omega_2} - 1\right)^2}{x_0}, \nonumber \\
    \alpha_{33} &= 16 J_{12}^2 \left[\gamma_A + \gamma_B \coth\left(\frac{\beta\omega_2}{2}\right)\right]\left[\gamma_A^2\left(e^{\beta\omega_2} - 1\right)^2 - e^{\beta\omega_2}\left(e^{\beta\omega_2} + 1\right)\gamma_B^2\right], \nonumber \\
    \eta_{33} &= \eta_{00}\left(e^{\beta\omega_2} + 1\right),
    \label{qtr_matrix_elements}
\end{align}

\end{widetext}
with $x_0 = \gamma_B - \gamma_A + e^{\beta\omega_2}\left(\gamma_A + \gamma_B\right)$. 

\section{Details on the forms of heat currents and proof of quantum thermal version of Kirchhoff's current law}

Consider a general system of $n$ nodes (labeled $1$, $2$, $3$, ..., $n$), where each node is coupled to all other nodes. Further, each node is weakly coupled to its respective bosonic thermal bath (labeled $I$, $II$, $III$, ..., $N$). The system's Hamiltonian in this setup is given by
\begin{align}
    H_S = \sum_{k = 1}^n H_k + \sum_{l,k=1, l<k}^n H_{lk},
\end{align}
where $H_k$ is the Hamiltonian of the node (here, it is $H_k = \frac{\omega_k}{2}\sigma^z_k$) and $H_{lk}$ is the coupling Hamiltonian between the nodes (here, we take $H_{lk} = J_{lk}\left(\sigma^x_l\sigma^x_k + \sigma^y_l\sigma^y_k\right)$ with $J_{lk}$ being the coupling strength). Under the Born-Markov and rotating wave approximations, the dynamics of the system (depicted by $\rho$) is given by the GKSL master equation 
\begin{align}
    \frac{d\rho}{dt} = -i\left[H_S, \rho\right] + \mathcal{D}_{I1}(\rho) + \mathcal{D}_{II2}(\rho) + ... + \mathcal{D}_{Nn}(\rho),
    \label{eq_drho_dt_current_form}
\end{align}
where $\mathcal{D}_{jk}(\rho) = \gamma_j(\tilde N_{jk} + 1)\left(L_k\rho L_k^\dagger - \frac{1}{2}\left\{L_k^\dagger L_k, \rho\right\}\right) + \gamma_j \tilde N_{jk}\left(L_k^\dagger\rho L_k - \frac{1}{2}\left\{L_k L_k^\dagger, \rho\right\}\right)$, and $\tilde N_{jk} = \frac{1}{e^{\beta_j\omega_k} - 1}$, with $\beta_j = T_j^{-1}$ for $(j, k) = (I, 1),  (II, 2), ..., (N, n)$. Further, $L_k$ are the Lindblad jump operators, which in the case of qubits are $\sigma^-_k$. Let us pick a node $m$ (Hamiltonian $H_m$) from the $n$ nodes. This node $m$ would have quantum heat currents from all the sources, that is, from all the other nodes that interact with it, as well as from the bosonic bath it is in interaction with (let the corresponding bath be $M$). The time derivative of the expectation value of $H_m$ is 
\begin{align}
    \frac{d}{dt}{\rm Tr}\left[H_m\rho\right] = {\rm Tr}\left[\frac{d H_m}{dt}\rho\right] + {\rm Tr}\left[H_m\frac{d \rho}{dt}\right], 
\end{align}
where the left-hand side of the above equation denotes the rate of change in the energy of the node $m$. On the right-hand side (RHS), the first term is the power, and the second term denotes the net heat current directed towards node $m$. Note that, in the present scenario, the Hamiltonian $H_m$ is constant; therefore, the first term on the RHS of the above equation becomes zero. Thus, $\frac{d}{dt}{\rm Tr}\left[H_m\rho\right] =  {\rm Tr}\left[H_m\frac{d \rho}{dt}\right]$. Now, we multiply Eq.~\eqref{eq_drho_dt_current_form} with $H_m$ as 
\begin{align}
    H_m \frac{d\rho}{dt} &= -i H_m\left[H_S, \rho\right] + H_m\mathcal{D}_{I1}(\rho) + H_m\mathcal{D}_{II2}(\rho) \nonumber \\
    &+ ... + H_m\mathcal{D}_{Nn}(\rho).
\end{align}
Taking the trace on both sides in the above equation, we get
\begin{align}
    {\rm Tr}\left[H_m\frac{d \rho}{dt}\right] &= -i {\rm Tr}\left[H_m H_S\rho - H_m\rho H_S\right] + {\rm Tr}\left[H_m\mathcal{D}_{I1}(\rho)\right] \nonumber\\
    &+ {\rm Tr}\left[H_m\mathcal{D}_{II2}(\rho)\right] + ... + {\rm Tr}\left[H_m\mathcal{D}_{Nn}(\rho)\right].
    \label{eq_net_current_rho_1}
\end{align}
It can be easily verified from the above equation that ${\rm Tr}\left[H_m\mathcal{D}_{jk}(\rho)\right] = 0$ for $(j, k) = (I, 1),  (II, 2), ..., (N, n)$ except when $j = M$ and $k = m$. Therefore, Eq.~\eqref{eq_net_current_rho_1} reduces to 
\begin{align}
    {\rm Tr}\left[H_m\frac{d \rho}{dt}\right] &= -i {\rm Tr}\left[H_m H_S\rho - H_m\rho H_S\right] + {\rm Tr}\left[H_m\mathcal{D}_{Mm}(\rho)\right].
    \label{eq_net_current_rho_2}
\end{align}
Further, the first term on the RHS in the above equation can be rewritten as 
\begin{align}
    -i {\rm Tr}\left[H_m H_S\rho - H_m\rho H_S\right] &= -i {\rm Tr}\left[H_m H_S\rho - H_SH_m\rho\right] \nonumber \\
    &= -i {\rm Tr}\left(\left[H_m, H_S\right]\rho\right).
\end{align}
The commutator of $H_m$ with $H_S$ can further be simplified to give $[H_m, H_S] = -\sum_{l = 1, l\ne m}^n[H_{lm}, H_{m}]$, using which we can write
\begin{align}
    -i {\rm Tr}\left[H_m H_S\rho - H_m\rho H_S\right] = \sum_{l = 1, l\ne m}^n i{\rm Tr}\left(\left[H_{lm}, H_{m}\right]\rho\right). 
    \label{eq_current_qubit_qubit}
\end{align}
Substituting the above equation in Eq.~\eqref{eq_net_current_rho_2}, we get the net current directed towards node $m$ as 
\begin{align}
    {\rm Tr}\left[H_m\frac{d \rho}{dt}\right] = \sum_{l = 1, l\ne m}^n \left\{i{\rm Tr}\left(\left[H_{lm}, H_{m}\right]\rho\right)\right\} + {\rm Tr}\left[H_m\mathcal{D}_{Mm}(\rho)\right].
    \label{eq_final_net_current}
\end{align}
It can be observed from the above equation that the heat current $\mathcal{J}_{lm}$ directed towards node $m$ from any other node $l$ is given by
\begin{align}
    \mathcal{J}_{lm} = i{\rm Tr}\left(\left[H_{lm}, H_{m}\right]\rho\right),
    \label{heat_current_in_between_qubits}
\end{align}
and the heat current $\mathcal{J}_{Mm}$ from bath $M$ (attached to node $m$) to node $m$ is given by 
\begin{align}
    \mathcal{J}_{Mm} = {\rm Tr}\left[H_m\mathcal{D}_{Mm}(\rho)\right].
    \label{eq_current_from_bath}
\end{align}

Further, in the steady state condition, the net current directed towards node $m$, Eq.~\eqref{eq_final_net_current}, becomes zero, that is, ${\rm Tr}\left[H_m\frac{d \rho^{SS}}{dt}\right] = 0$. This implies,
\begin{align}
    \sum_{l = 1, l\ne m}^n \mathcal{J}_{lm} + \mathcal{J}_{Mm} = 0.
\end{align}
The above equation proves the quantum thermal version of Kirchhoff's current law.
\begin{figure*}
    \centering
    \includegraphics[width=0.75\linewidth]{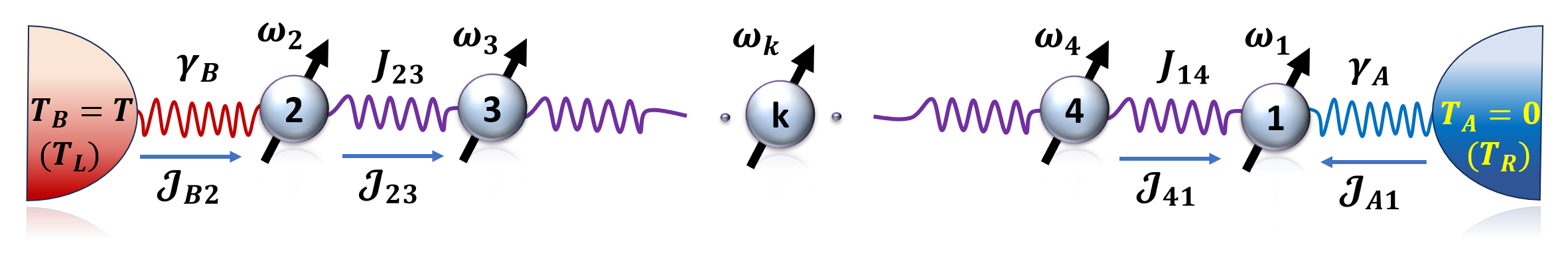}
    \caption{A schematic diagram of a spin-chain (containing $n$ spins) connected to two baths at different temperatures.}
    \label{fig_comparison_with_the_established_forms}
\end{figure*}

\subsection{Comparison of the derived heat current with the established forms in the literature}
We derived the heat current via qubit $m$, as discussed above, by considering it as a node, receiving heat from the local bath it is attached to, and from the neighboring spins. The form of the heat current the qubit receives from the local bath $M$, ${\rm Tr}\left[H_m\mathcal{D}_{Mm}(\rho)\right]$ matches with the energy current from the site-resolved continuity equation discussed in~\cite{Mendoza-Arenas_2013, lendi}. Further, the form of the heat current that qubit $m$ receives from other qubits $i{\rm Tr}\left(\left[H_{lm}, H_{m}\right]\rho\right)$ is similar to the one obtained in~\cite{Tanimura_2016}, where it was called the system heat current. There, the system interacts with $k$ baths and the heat current from the $k^{th}$ bath is given by $i{\rm Tr}\left(\left[H_{km}, H_{m}\right]\rho\right)$ ($H_{km}$ is the interaction Hamiltonian between the system and the $k^{th}$ bath). This relationship was shown to hold even in the strong coupling regime. Here, the qubit $m$ can be considered as the system, while the qubits it interacts with would be analogous to a bath supplying the heat current, such as $i{\rm Tr}\left(\left[H_{lm}, H_{m}\right]\rho\right)$, from the $l^{th}$ qubit.

Further, the total current $\mathcal{I}$ entering the system from the various baths it is connected to can be found using the time derivative of the expectation value $\braket{H_S}$, where $H_S$ is the system Hamiltonian composed of all qubits. This can be found using the system-resolved continuity equations and has the following form~\cite{lendi},
\begin{align}
 \mathcal{I} = \frac{d}{dt}\braket{H_S} = \sum_k \mathcal{I}_k =  \sum_{k}{\rm Tr}\left[H_S \mathcal{D}_k(\rho)\right],
 \label{total_current}
\end{align}
where $\mathcal{D}_k$ has the similar form as discussed in~\eqref{eq_drho_dt_current_form}. In the steady state condition, the sum of all such currents becomes zero. Particularly, in the case of a spin chain connected to a right [with dissipator $\mathcal{D}_R(\rho)$] and a left bath [with dissipator $\mathcal{D}_L(\rho)$] at different temperatures, we get $\mathcal{I} = \mathcal{I}_L + \mathcal{I}_R = 0$, which implies
\begin{align}
    \mathcal{I}_L = -\mathcal{I}_R.
\end{align}
This form shows that the current $\mathcal{I}_L$ entering the system from the left bath is equal to the current $\mathcal{I}_R$ entering the right bath and is useful in discussions of thermal conductivity inside a spin chain, see Fig.~\ref{fig_comparison_with_the_established_forms}, connected to two reservoirs at different temperatures~\cite {lendi, Dhar2008, Dhar_Saito_Hanggi, Junaid_Abhishek}. Note that the form of current derived in Eq.~\eqref{eq_current_from_bath} can be different from the current $\mathcal{I}_{L(R)}$. This is so because the focus of this work is to find the thermal transport via a given node (qubit). Interestingly, however, in the scenario of a spin chain connected to two baths at different temperatures at their extreme ends, the current $\mathcal{I}_k$ and the current $\mathcal{J}_{Mm}$ become the same for equal frequencies $\omega_j$ of the qubits in the spin-chain. 
\begin{figure}
    \centering
    \includegraphics[width=1\linewidth]{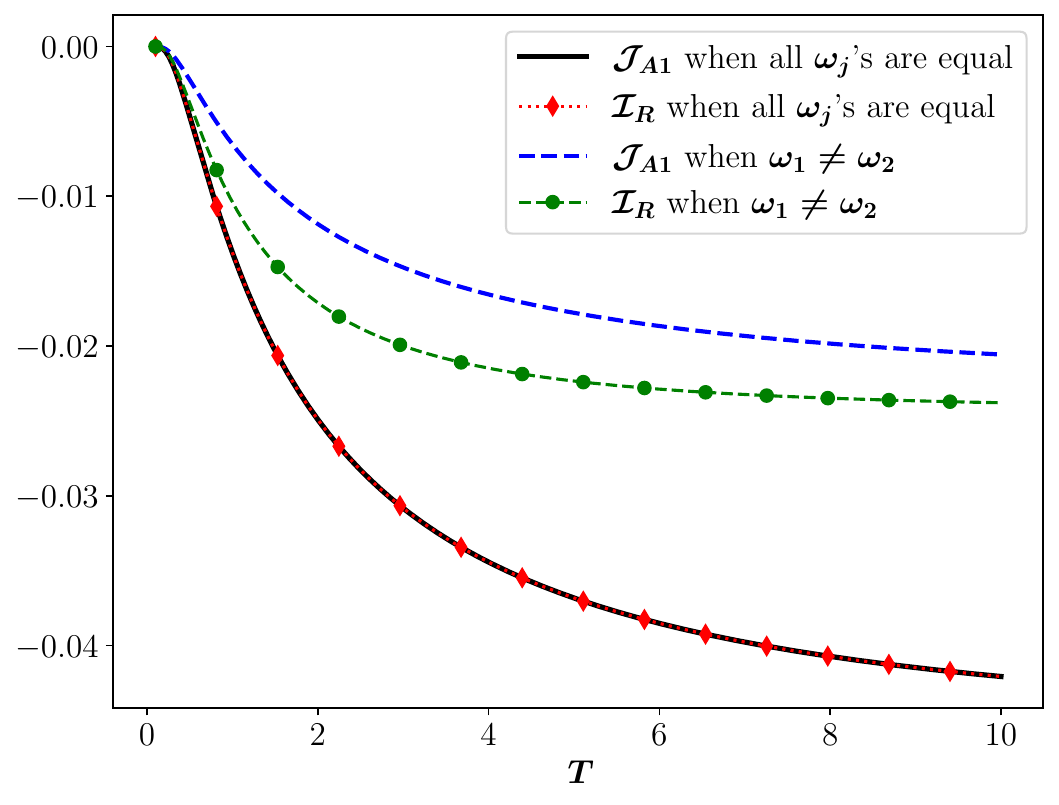}
    \caption{Variation of the currents $\mathcal{J}_{A1}$ from bath $A$ to qubit 1, and $\mathcal{I}_R$ from right bath to the system with temperature $T$ of bath $B$, see Fig.~\ref{fig_comparison_with_the_established_forms}. The parameters are $n = 4, J_{14} = 0.65, J_{23} = 0.56, J_{34} = 0.75, \gamma_A = 0.1$, and $\gamma_B = 0.05$.}
    \label{fig_comparison_current_and_total_current}
\end{figure}
This is illustrated in Fig.~\ref{fig_comparison_current_and_total_current}. 
In fact, the currents $\mathcal{I}_{L(R)}$, $\mathcal{J}_{A1}$, and $\mathcal{J}_{B2}$ are related by
\begin{align}
    \mathcal{I}_R &= \mathcal{J}_{A1} + {\rm Tr}\left[H_{1n}\mathcal{D}_{A1}\left(\rho\right)\right], \nonumber \\
    \mathcal{I}_L &= \mathcal{J}_{B2} + {\rm Tr}\left[H_{23}\mathcal{D}_{B2}\left(\rho\right)\right],
\end{align}
where $H_{1n}$ and $H_{23}$ are the coupling Hamiltonians between qubits 1 and $n$ and between qubits 2 and 3, respectively.
Hence, the formalism developed in this work can also be adapted to studies in thermal conductivity.

In the literature, we also find a description of a spin current~\cite{Poletti_2018, Poulsne_2022, Poulsen_2024, Mendoza_2024} (or magnetization current) for the spin-chain models, given by
\begin{align}
    I_M = 2 J \left\langle\sigma^x_j\sigma^y_{j+1} - \sigma^y_{j}\sigma^x_{j+1}\right\rangle,
    \label{eq_spin_current}
\end{align}
where $J$ is the qubit-qubit coupling strength. This current is site-independent since the current entering a given site is the same as the current leaving the other site. 
This can be derived using the spin continuity equation. Consider a spin at site $k$ in the Heisenberg spin chain model; see Fig.~\ref{fig_comparison_with_the_established_forms}. The spin continuity equation can be written as 
\begin{align}
    \frac{d\braket{\sigma^z_k}}{dt} = (j_{k-1, k} - j_{k, k+1}),
    \label{eq_spin_continuity_eq}
\end{align}
where $j_{k-1, k}$ is the flux incoming from the $k-1$-th qubit to $k$-th qubit and similarly $j_{k, k+1}$ is the flux outgoing from the $k$-th qubit to $k+1$-th qubit. 
The left-hand side in the above equation can be obtained using the master equation~\eqref{eq_drho_dt_current_form} as 
\begin{align}
    \frac{d\braket{\sigma^z_k}}{dt} = i\braket{[H_S, \sigma^z_k]} + {\rm Tr}\left[\sigma^z_k\mathcal{D}_{A1}\right] + {\rm Tr}\left[\sigma^z_k\mathcal{D}_{B2}\right],
\end{align}
where the terms ${\rm Tr}\left[\sigma^z_k\mathcal{D}_{A1}\right]$ and  ${\rm Tr}\left[\sigma^z_k\mathcal{D}_{B2}\right]$ become zero for $k\ne 1, 2$. Further, the commutator $[H_S, \sigma^z_k]$ simplifies to $[H_S, \sigma^z_k] = [H_{k-1, k}, \sigma^z_k] + [H_{k, k+1}, \sigma^z_k]$, where $H_{k-1,k}$ and $H_{k,k+1}$ are the interaction Hamiltonians between qubits at sites $k-1$ and $k$ and between qubits at sites $k$ and $k+1$, respectively. This follows from the fact that all other terms in $H_S$ commute with $\sigma^z_k$. Now, the above equation can be rewritten as
\begin{align}
    \frac{d\braket{\sigma^z_k}}{dt} = i\left(\left\langle[H_{k-1, k}, \sigma^z_k]\right\rangle + \left\langle[H_{k, k+1}, \sigma^z_k]\right\rangle\right).
\end{align}
On comparing the above equation with the spin continuity equation~\eqref{eq_spin_continuity_eq}, we get the form of spin current $j_{k, k+1}$ from qubit $k$ to $k+1$ to be
\begin{align}
    j_{k, k+1} = -i\left\langle\left[H_{k, k+1}, \sigma^z_k\right]\right\rangle,
\end{align}
which, using the form of the interaction Hamiltonian $H_{k, k+1} = J\left(\sigma^x_k\sigma^x_{k+1} + \sigma^y_k\sigma^y_{k+1}\right)$ (Heisenberg XX interaction), becomes the spin current [Eq.~\eqref{eq_spin_current}]
\begin{align}
    j_{k, k+1} = 2J\left\langle\sigma^x_k\sigma^y_{k+1} - \sigma^y_k\sigma^x_{k+1}\right\rangle = I_M.
\end{align}%

A form of heat current is also found by multiplying the spin current by the frequency of the qubit~\cite{Poulsen_2024}. 
However, on comparing the spin current $I_M$ and the form of heat current obtained in Eq.~\eqref{heat_current_in_between_qubits}, we find that the magnitudes of both are related by 
\begin{align}
    |\mathcal{J}_{lk}| = \frac{\omega_k}2 |I_M|.
\end{align}
Here, the factor $\omega_k/2$ is important. Thus, for example, in the case of the node connected to the bath, say, qubit $1$ in Fig.~\ref{fig_comparison_with_the_established_forms}, the quantum thermal version of Kirchhoff's current law at node 1 will only be satisfied if the factor $\omega_1/2$ is present. That is the sum of heat current $\mathcal{J}_{A1} + \mathcal{J}_{41} = \mathcal{J}_{A1} + \frac{\omega_1}{2}I_M = 0$. The heat currents derived in this work naturally take care of different scenarios, whether the heat current is calculated at the node connected to the bath or in between the nodes. In contrast, the spin current pertaining to the spin interacting with the bath is calculated by invoking boundary conditions~\cite{spin_current_landi}.

The forms of heat current, Eqs.~\eqref{heat_current_in_between_qubits} and~\eqref{eq_current_from_bath}, are more general in the sense that the expressions of heat current, along with their correct frequency scaling factors, can be obtained in scenarios where there are chains of harmonic oscillators or a mixed chain of harmonic oscillators and spins, and for arbitrary interactions between them.
Further, there is an advantage in using these forms; for example, in a complicated scenario, such as a circuit of a thermal super Wheatstone bridge, the magnitude and direction of the heat current can be exactly determined.

\section{Quantum Thermal Transformer}
Here, we provide proof of the expression regarding the quantum thermal transformer. Consider the quantum thermal current between two qubits $\mathcal{J}_{jk} =
i{\rm Tr}\left\{\rho^{SS}[H_{jk}, H_k]\right\}$, where $H_k = \frac{\omega_k}{2}\sigma^z_k$ and $H_{jk} = J_{jk}\left(\sigma^x_j\sigma^x_k + \sigma^y_j\sigma^y_k\right),$ ($j\ne k$). Now, 
\begin{align}
[H_{jk}, H_j]&= H_{jk} H_j  - H_j H_{jk}  \nonumber \\
&= J_{jk}\frac{\omega_j}{2} \left( \sigma^x_j \sigma^x_k + \sigma^y_j \sigma^y_k \right)  \sigma^z_j - J_{jk} \frac{\omega_j}{2} \sigma^z_j \left( \sigma^x_j \sigma^x_k + \sigma^y_j \sigma^y_k \right)\nonumber \\
&= iJ_{jk} \omega_j \left[ \sigma^x_j \sigma^y_k - \sigma^y_j \sigma^x_k \right], ~~~~ \text{and} \nonumber \\
[H_{jk}, H_k]&= H_{jk} H_k  - H_k H_{jk}  \nonumber \\
&= J_{jk}\frac{\omega_k}{2} \left( \sigma^x_j \sigma^x_k + \sigma^y_j \sigma^y_k \right)  \sigma^z_k - J_{jk} \frac{\omega_j}{2} \sigma^z_k \left( \sigma^x_j \sigma^x_k + \sigma^y_j \sigma^y_k \right)\nonumber \\
&= iJ_{jk} \omega_k \left[ \sigma^y_j \sigma^x_k - \sigma^x_j \sigma^y_k \right],
\label{inter_qubit_current_pauli_form}
\end{align}
From the above, we can verify that 
\begin{align}
    \frac{[H_{jk}, H_j]}{\omega_j} = - \frac{[H_{jk}, H_k]}{\omega_k}.
\end{align}
Therefore, we get a relationship between the currents between two qubits $\mathcal{J}_{jk}$ (taking qubit $k$ as the node) and $\mathcal{J}_{kj}$ (taking qubit $j$ as the node) as 
\begin{align}
    \frac{\mathcal{J}_{jk}}{\omega_k} = -\frac{\mathcal{J}_{kj}}{\omega_j}.
\end{align}

\section{Thermal potentials}
Consider a quantum thermal circuit made up of three qubits. The Hamiltonian for this three-qubit setup is given by
\begin{align}
    \widetilde H_S &= H_1 + H_2 + H_3 + H_{12} + H_{13} + H_{23}, \nonumber \\
    &= \frac{\omega_1}{2}\sigma_1^z + \frac{\omega_2}{2}\sigma_2^z + + \frac{\omega_3}{2}\sigma_3^z + J_{12}\left(\sigma^x_1\sigma^x_2 + \sigma^y_1\sigma^y_2\right) \nonumber \\
    &+ J_{13}\left(\sigma^x_1\sigma^x_3 + \sigma^y_1\sigma^y_3\right) + J_{23}\left(\sigma^x_2\sigma^x_3 + \sigma^y_2\sigma^y_3\right),
\end{align}
where qubits $l$ and $k$ interact with each other via Heisenberg $XX$ type interaction with the interaction strength $J_{lk}$ and $\omega_k$ being the transition frequency of the $k$-th qubit. Under the Born-Markov and secular approximations, Eq. (\ref{appendix_qtr_master_eq}) with $\widetilde H_S$ in place of $H_S$ dictates the dynamics of the three-qubit system. By equating this to zero, we find the steady-state $\tilde \rho^{SS}$ of the three-qubit system. The effective temperatures of the qubits 1, 2, and 3 ($T_1$, $T_2$, and $T_3$, respectively), using the steady-state $\tilde \rho^{SS}$ of the three-qubit system, are given by
\begin{align}
    T_1 &= \frac{\omega_1}{\log\left(\frac{1}{\tilde \rho^{SS}_{00} + \tilde \rho^{SS}_{11} + \tilde \rho^{SS}_{22} + \tilde \rho^{SS}_{33}} - 1\right)}, &&
    T_2 = \frac{\omega_2}{\log\left(\frac{1}{\tilde \rho^{SS}_{00} + \tilde \rho^{SS}_{11} + \tilde \rho^{SS}_{44} + \tilde \rho^{SS}_{55}} - 1\right)}, \nonumber \\
    T_3 &= \frac{\omega_3}{\log\left(\frac{1}{\tilde \rho^{SS}_{00} + \tilde \rho^{SS}_{22} + \tilde \rho^{SS}_{44} + \tilde \rho^{SS}_{66}} - 1\right)}.
    \label{eq_three_qubit_effective_temp}
\end{align}%
The corresponding thermal potential differences are given by 
\begin{align}
    V_{1A} &= T_1 - 0; && V_{21} = T_2 - T_1; && V_{32} = T_3 - T_2; \nonumber \\
    V_{13} &= T_1 - T_3; && V_{B2} = T - T_2.
\end{align}
In terms of the elements of the steady state $\tilde \rho^{SS}$ of the three-qubit circuit, the quantum thermal currents between qubits 1 and 3 and between qubits 1 and 2, taking qubit 1 as a node, are given by 
\begin{align}
    \mathcal{J}_{31} &= -4J_{13}\omega_1\left[\Im{(\tilde \rho^{SS}_{14})} + \Im{(\tilde \rho^{SS}_{36})}\right], \nonumber \\
    \mathcal{J}_{21} &= -4J_{12}\omega_{1}\left[\Im{(\tilde \rho^{SS}_{24})} + \Im{(\tilde \rho^{SS}_{35})}\right],
    \label{eq_three_qubit_q13_q12}
\end{align}
where $\Im(z)$ denotes imaginary part of $z$. Now, by defining a function 
\begin{align}
\tilde f_{13} &= \frac{1}{\left[\Im{(\tilde \rho^{SS}_{14})} + \Im{(\tilde \rho^{SS}_{36})}\right]}\left[\frac{\omega_3}{\omega_1\log\left(\frac{1}{\tilde \rho^{SS}_{00} + \tilde \rho^{SS}_{22} + \tilde \rho^{SS}_{44} + \tilde \rho^{SS}_{66}} - 1\right)} \right. \nonumber \\
&-\left. \frac{1}{\log\left(\frac{1}{\tilde \rho^{SS}_{00} + \tilde \rho^{SS}_{11} + \tilde \rho^{SS}_{22} + \tilde \rho^{SS}_{33}} - 1\right)}\right],
\label{eq_three_qubit_f13}
\end{align}
which is a function of $\omega_i, T_i, J_{ij}, T, \gamma_i$ (for $i, j = 1, 2, 3$), a relation between the quantum thermal heat current $\mathcal{J}_{31}$ and the thermal potential $V_{13}$ is given as 
\begin{align}
    \mathcal{J}_{31} = \frac{4J_{13}}{\tilde f_{13}}V_{13}.
\end{align}
This is Eq.~(7) in the main text. 
Further, using the relation $\frac{\mathcal{J}_{31}}{\omega_1} = -\frac{\mathcal{J}_{13}}{\omega_3}$, we can write
\begin{align}
    \mathcal{J}_{13} = -\frac{4\omega_3 J_{13}}{\omega_1\tilde f_{13}}V_{13}.
\end{align}
Similarly, we can find out various functions, such as $\tilde f_{12}$, and $\tilde f_{23}$, which can be used to identify the corresponding relation of the quantum thermal currents $\mathcal{J}_{12}, \mathcal{J}_{21}, \mathcal{J}_{23}$, and $\mathcal{J}_{32}$ with the quantum thermal potentials $V_{12}$ and $V_{23}$. 
\section{Quantum Thermal Diode}
Here, we apply the framework developed to a few other quantum thermal circuits, for example, the quantum thermal diode~\cite{circuit4, thermal_diode}, followed by the quantum thermal transistor. The Hamiltonian of the quantum thermal diode is similar to the quantum thermal resistor discussed above, Eq.~\eqref{qtr_ham}. The difference is that here, we take the temperature of bath $A$ to be non-zero. Accordingly, the dynamical master equation, Eq.~\eqref{appendix_qtr_master_eq}, changes to accommodate the factor $N_{th, A} = \frac{1}{e^{\omega_1/T_A} - 1}$ with the Lindblad operator $\sigma^+_1$. The steady-state for this system can be found similarly. We calculate the thermal heat current that flows between bath $A$ and qubit 1. Based on the temperature difference between baths $A$ and $B$, we determine the reverse and forward bias of the thermal diode. The voltage and, correspondingly, the thermal heat currents are zero when the temperatures of $A$ and $B$ are the same.
\begin{figure}
    \centering
    \includegraphics[width=1\columnwidth]{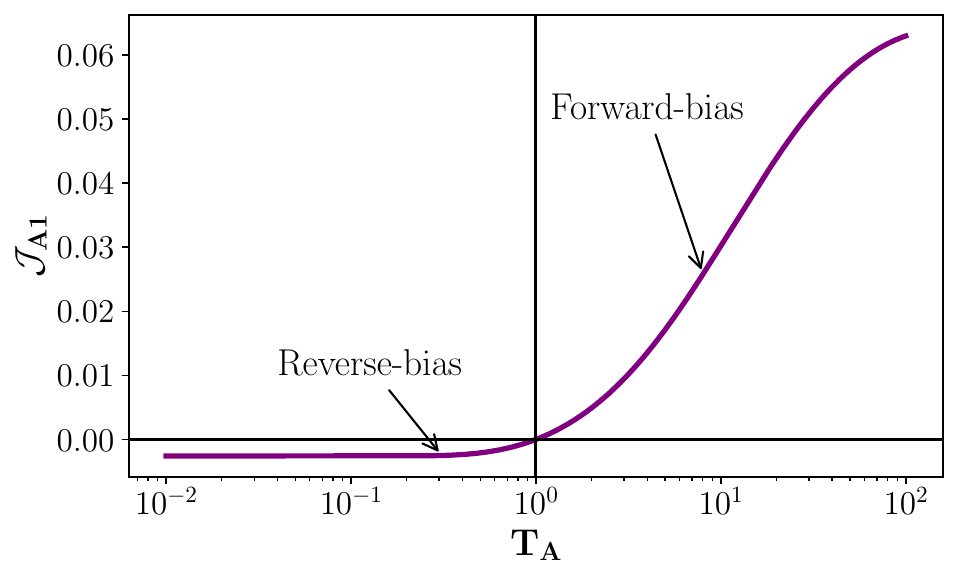}
    \caption{Variation of heat current $\mathcal{J}_{A1}$ from bath $A$ to qubit 1 as a function of temperature $T_A$ of bath $A$ for a quantum thermal diode. The parameters are: $\omega_1 = 1.5, \omega_2=1.5, J_{12} = 0.5, \gamma_A = 0.01, \gamma_B = 0.1, T_B = 1.0$.}
    \label{fig_quantum_thermal_diode}
\end{figure}
In Fig.~\ref{fig_quantum_thermal_diode}, we show the variation of quantum thermal heat current $\mathcal{J}_{A1}$ with the temperature of bath $A$. The forward and the reverse bias regions are specified explicitly in the figure. In the forward bias region, the heat current flows from bath $A$ to the qubit 1, whereas in the reverse bias region, the heat current flows from the qubit 1 to the bath $A$. Further, the heat current is zero when the temperatures $T_A$ and $T_B$ of the baths $A$ and $B$, respectively, are equal.

\section{Quantum Thermal Transistor}
The quantum thermal transistor circuit is an upgrade to the quantum thermal diode~\cite{circuit5, circuit6}. It also acts as an amplifier of the heat current. The circuit model requires three qubits and three baths.
\begin{figure}
    \centering
    \includegraphics[width=0.85\columnwidth]{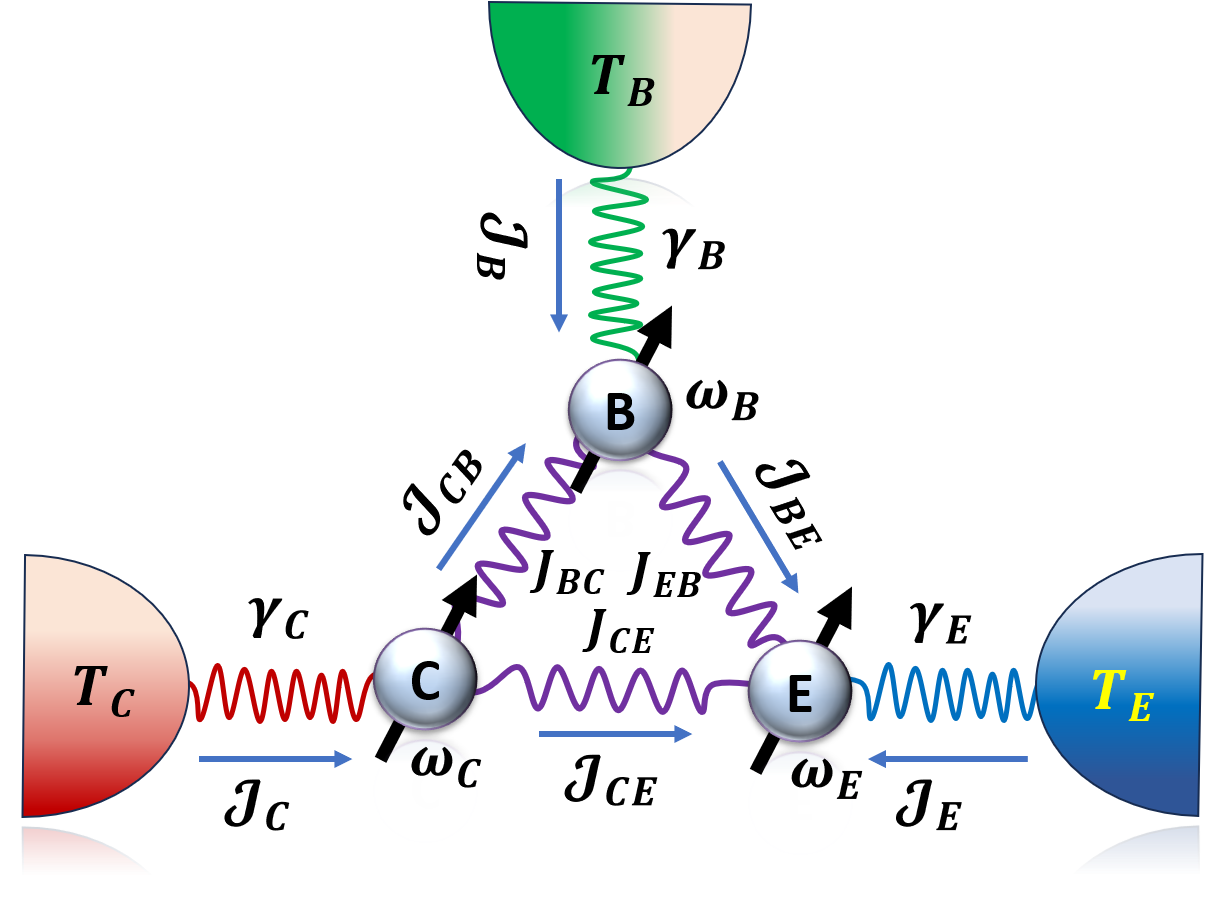}
    \caption{A schematic diagram to show the arrangement of qubits and the respective baths to function as a quantum thermal transistor. The labels $B$, $C$, and $E$ correspond to base, collector, and emitter, respectively.}
    \label{fig_quantum_thermal_transistor_schematic}
\end{figure}
A schematic diagram of the circuit is provided in Fig.~\ref{fig_quantum_thermal_transistor_schematic}. The labels $B$, $C$, and $E$ correspond to base, collector, and emitter, respectively. 
\begin{figure}
    \centering
    \includegraphics[width=1\linewidth]{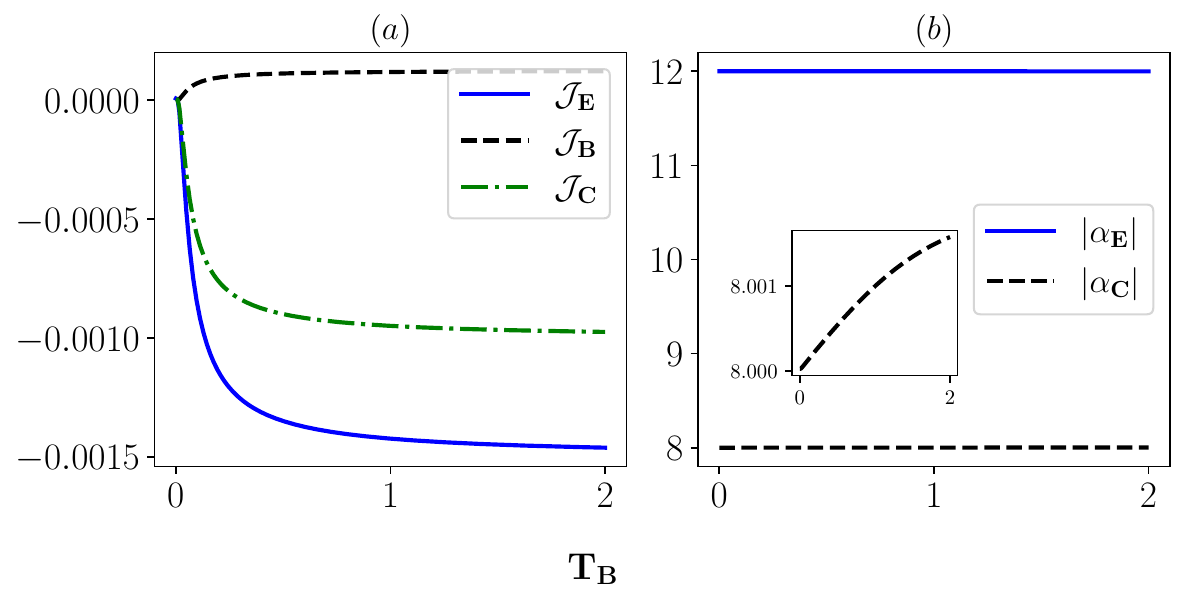}
    \caption{Variation of the (a) collector ($\mathcal{J}_C$), base ($\mathcal{J}_B$), and emitter ($\mathcal{J}_E$) thermal currents and (b) the corresponding amplification factor $\alpha_C$ and $\alpha_E$ as a function of the base temperature $T_B$. The parameters take the following values: $\omega_1 = \omega_2 = 1.0, \omega_3 = 0.05\omega_1, J_{12} = J_{13} = J_{23} = 1.0, T_E = T_C = 0.2, \gamma_E = 0.003, \gamma_B = 0.01, \gamma_C = 0.002$.}
    \label{fig_quantum_thermal_transistor}
\end{figure}%
Further, the three baths have non-zero temperatures. The interaction strength factors between the qubits $B$, $C$, and $E$ are $J_{BC}$, $J_{EB}$, and $J_{CE}$. The Hamiltonian of the system is given by
\begin{align}
    H_{trans} &= \frac{\omega_C}{2}\sigma^z_C + \frac{\omega_B}{2}\sigma^z_B + \frac{\omega_E}{2}\sigma^z_E + J_{BC}\left(\sigma^x_B\sigma^x_C + \sigma^y_B\sigma^y_C\right) \nonumber \\
    &+ J_{EB}\left(\sigma^x_E\sigma^x_B + \sigma^y_E\sigma^y_B\right) + J_{CE}\left(\sigma^x_C\sigma^x_E + \sigma^y_C\sigma^y_E\right).
\end{align}
The above Hamiltonian is along similar lines to our adaptation of thermal resistors, diodes, and other thermal circuits discussed here. 
The master equation dictating the dynamics of the system is given by 
\begin{align}
    \frac{d\rho}{dt} &= -i[H_{trans}, \rho] + \sum_{k = B, C, E}\left(\gamma_k\left(N_{th, k} + 1\right)\left[\sigma^-_k\rho\sigma^+_k - \frac{1}{2}\left\{\sigma^+_k\sigma^-_k\right\}\right] \right. \nonumber \\
    &+\left. \gamma_k N_{th, k}\left[\sigma^+_k\rho\sigma^-_k - \frac{1}{2}\left\{\sigma^-_k\sigma^+_k\right\}\right]\right),
\end{align}
where $N_{th, k} = \frac{1}{e^{\omega_k/T_k} - 1}$ for $k = B, C, E$. 
The base, emitter, and collector currents $\mathcal{J}_B$, $\mathcal{J}_E$, and $\mathcal{J}_C$, respectively, for the quantum thermal transistor are found using the steady-state $\rho^{SS}_{trans}$ in Eq.~\eqref{eq_current_from_bath} and are plotted in Fig.~\ref{fig_quantum_thermal_transistor}(a) as a function of the base temperature $T_B$. It can be observed that the base current only changes slightly, whereas, for the same change in temperature, the collector and emitter currents change rapidly. To quantify the amplifications in the collector and emitter currents, the factor 
\begin{align}
\alpha_P = \frac{\mathcal{J}_P}{\mathcal{J}_B},
\end{align}
for $P = C, E$ is plotted in Fig.~\ref{fig_quantum_thermal_transistor}(b). It can be observed that the collector and emitter currents get amplified by 8 and 12 times, respectively, for the given set of parameters. 

\section{Further details on quantum thermal super Wheatstone bridge}
\begin{figure}
    \centering
    \includegraphics[width=1\linewidth]{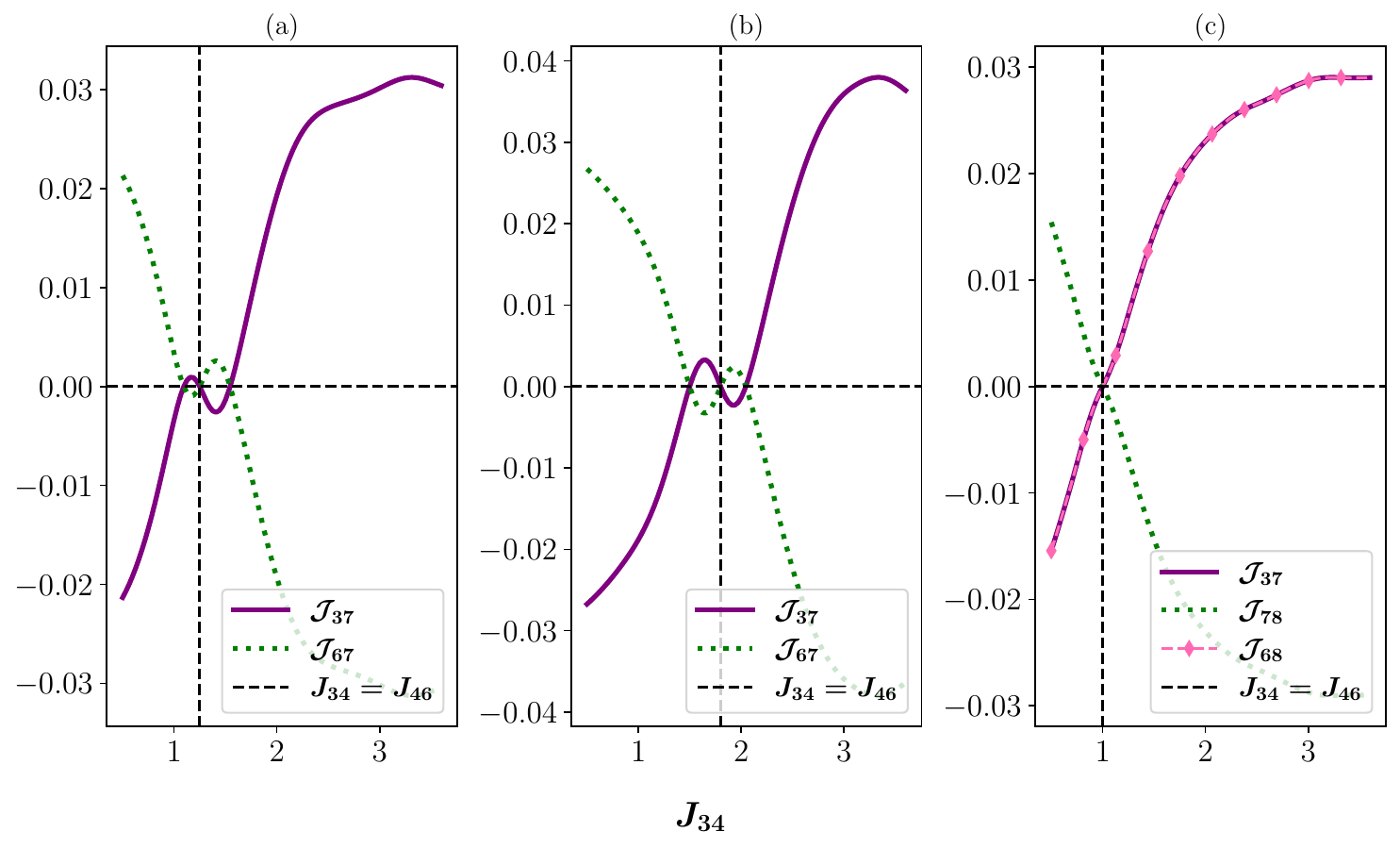}
    \caption{Variation of the thermal currents $\mathcal{J}_{37}$ and $\mathcal{J}_{67}$ in (a), (b) with the interaction strength $J_{34}$ between qubits 3 and 4 when condition 2 of the super Wheatstone bridge (Fig.~5 in the main text) is in effect ($J_{23} = J_{16} = J_{47} = J_{57} = 0$). The balance condition of this circuit is when $J_{34} = J_{46}$, $J_{35} = J_{56}$, and $\omega_3 = \omega_7 = \omega_6$. (a), (b) and (c) show the plot for different values of interaction strength $J_{46}$. Further, all other $J_{jk}$'s and $\omega_j$'s are taken randomly between 1 and 2; and $T =10$, $\gamma_A = 0.1$ and $\gamma_B = 0.05$. In (c), we plot the variation of heat currents $\mathcal{J}_{37}, \mathcal{J}_{78}$, and $\mathcal{J}_{68}$ with the interaction strength $J_{34}$ for the circuit shown in Fig.~\ref{fig_variations_in_quantum_thermal_super_Wheatstone_bridge}(a). The balance condition is: $J_{34} = J_{46}$, $J_{35} = J_{56}$, and $\omega_3 = \omega_7 = \omega_8 =\omega_6$.}
    \label{fig_super_Wheatstone_bridge_condition_2}
\end{figure}

\begin{figure}
    \centering
    \includegraphics[width=1\linewidth]{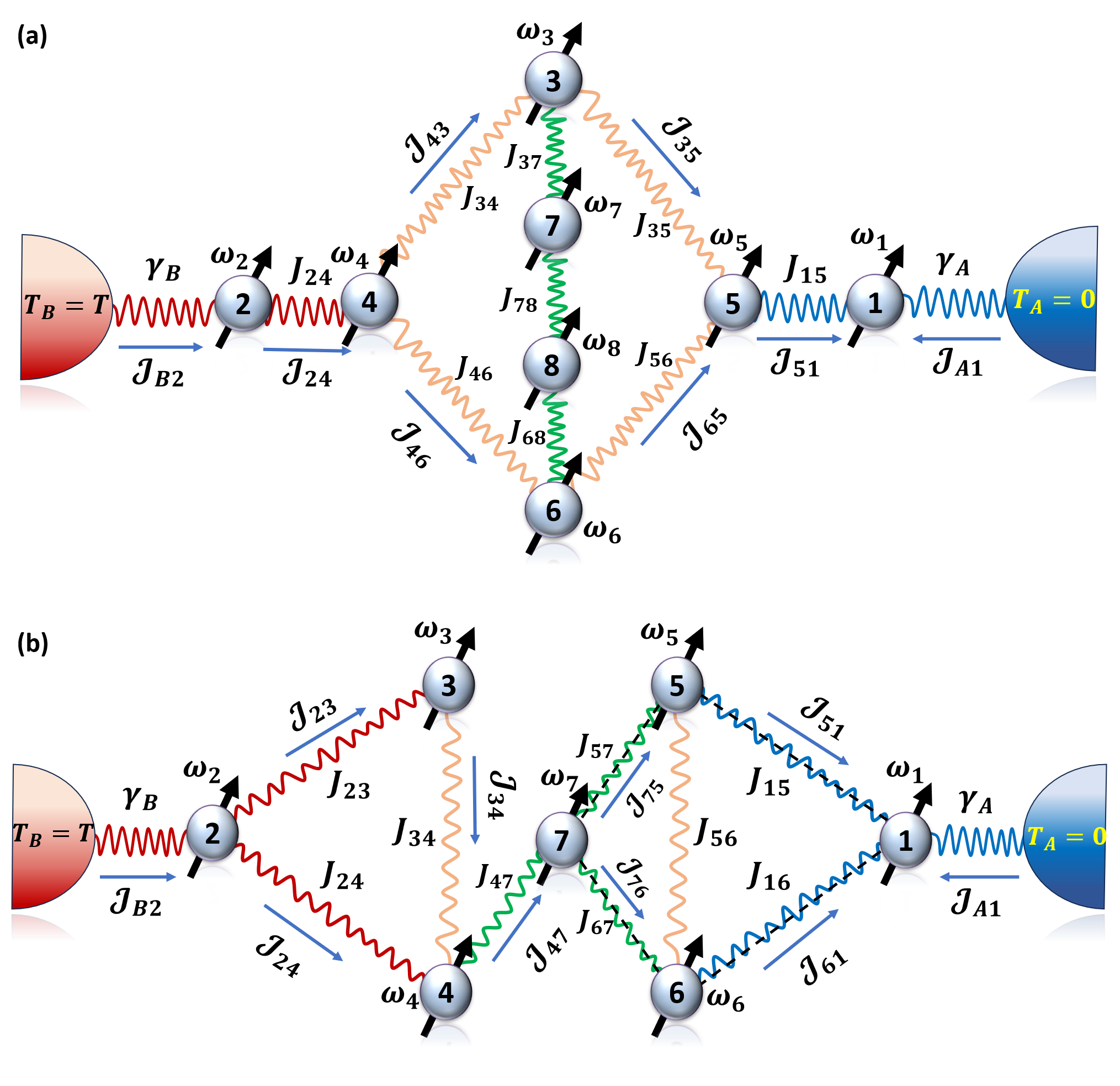}
    \caption{Schematic diagrams depicting (a) condition 2 ($J_{23} = J_{16} = J_{47} = J_{57} = 0$) and (b) condition 3 ($J_{35} = J_{37} = J_{46} = 0$) of the quantum thermal super Wheatstone bridge. In (a), we have taken an extra qubit 8, and (b) denotes a standard Wheatstone bridge between qubits 1, 5, 7, and 6.}
    \label{fig_variations_in_quantum_thermal_super_Wheatstone_bridge}
\end{figure}

Here, we elaborate on conditions 2 and 3 of the quantum thermal super Wheatstone bridge. First, we discuss condition 2 as depicted in Fig.~5 of the main manuscript, where qubit 7 remains unaffected by the thermal transport around it in the balanced condition. From Fig.~\ref{fig_super_Wheatstone_bridge_condition_2}(a) and (b), we can observe that there are other points, too, where the currents $\mathcal{J}_{37}$ and $\mathcal{J}_{67}$ go to zero. However, these points are absent in Fig.~6(b) of the main manuscript, which depicts zero current via qubit 7 at the balance condition $J_{34} = J_{46}$, $J_{35} = J_{56}$, and $\omega_3 = \omega_7 = \omega_6$. This shows that there may be points where the thermal current via qubit 7 may go to zero; however, they are always zero at the balance condition, and in this sense, we call this balance condition universal. In Fig.~\ref{fig_variations_in_quantum_thermal_super_Wheatstone_bridge}(a), we show an extended version of condition 2 of the super Wheatstone bridge.
Here, we want two central qubits, for example, qubits 7 and 8 in Fig.~\ref{fig_variations_in_quantum_thermal_super_Wheatstone_bridge}(a), to be unaffected by the heat transport around them. In other words, the heat currents $\mathcal{J}_{37}=\mathcal{J}_{78}=\mathcal{J}_{68}=0$. We observe, from Fig.~\ref{fig_super_Wheatstone_bridge_condition_2}(c), that under a similar balance condition: $J_{34} = J_{46}$, $J_{35} = J_{56}$, and $\omega_3 = \omega_7 = \omega_8 =\omega_6$, we get zero current around both the central qubits, that is, $\mathcal{J}_{37} = \mathcal{J}_{78} = \mathcal{J}_{68} = 0$. In this way, even if we add multiple central qubits, we will get zero current. This can be understood as a consequence of the quantum thermal version of Kirchhoff's first law as the net current coming to qubits 7 and 8 should be zero and since they are connected via a single wire running from qubit 3 to qubit 6, if the current in any part of the wire, for example, between qubits 3 and 7 ($\mathcal{J}_{37}$), is zero, it has to be zero for all the other parts, and hence all the central qubits will remain unaffected by the thermal transport around them if the balance condition is achieved. Note that the above balance condition still holds even if the interaction strengths $J_{37}, J_{78}$, and $J_{68}$ are not equal.

Condition 3 of the super Wheatstone bridge is depicted in Fig.~\ref{fig_variations_in_quantum_thermal_super_Wheatstone_bridge}(b), where the standard quantum thermal Wheatstone bridge is extracted as a subset of the super Wheatstone bridge, formed using qubits 1, 5, 7, and 6. A detailed analysis of this type of circuit is done below. Remarkably, many more of these types of standard Wheatstone bridges can be extracted from the super Wheatstone bridge (Fig.~\ref{fig_quantum_thermal_super_Wheatstone_bridge} of the main manuscript). Further, using the quantum thermal version of Kirchhoff's current law, we can derive the following relation between the current from qubit 4 to qubit 7, $\mathcal{J}_{47}$, and the current from bath $B$ to qubit 2, $\mathcal{J}_{B2}$,
\begin{align}
 \mathcal{J}_{47} = \frac{\omega_7}{\omega_2} \mathcal{J}_{B2}.   
\end{align}
This relation suggests that qubit 7 gets the heat current from bath $B$ in a manner analogous to a direct interaction up to a scaling factor of $\omega_7/\omega_2$, effectively forming a configuration resembling the standard Wheatstone bridge.
\subsection{The standard quantum thermal Wheatstone bridge}
The classical Wheatstone bridge is a device primarily used to determine an unknown resistance. In a circuit of four resistances, one is a tunable resistance, and the other is the unknown, whose value is determined when the ratio between them becomes equal to that of the known resistances. In the case of the quantum Wheatstone bridge, the proposed device would determine the unknown interaction strength of a Hamiltonian. A pictorial comparison between a classical Wheatstone bridge and a quantum one is depicted in~\cite{QWB}, wherein a model of a quantum Wheatstone bridge was proposed, with the balance condition determining the unknown coupling strength obtained by observing a drop in population of Bell states caused by controlling a tunable coupling. On the other hand, our approach to finding the balance condition is by manipulating the heat current through the circuit, which is in tune with the actual classical Wheatstone bridge. 
\begin{figure}
    \centering
    \includegraphics[width=0.95\linewidth]{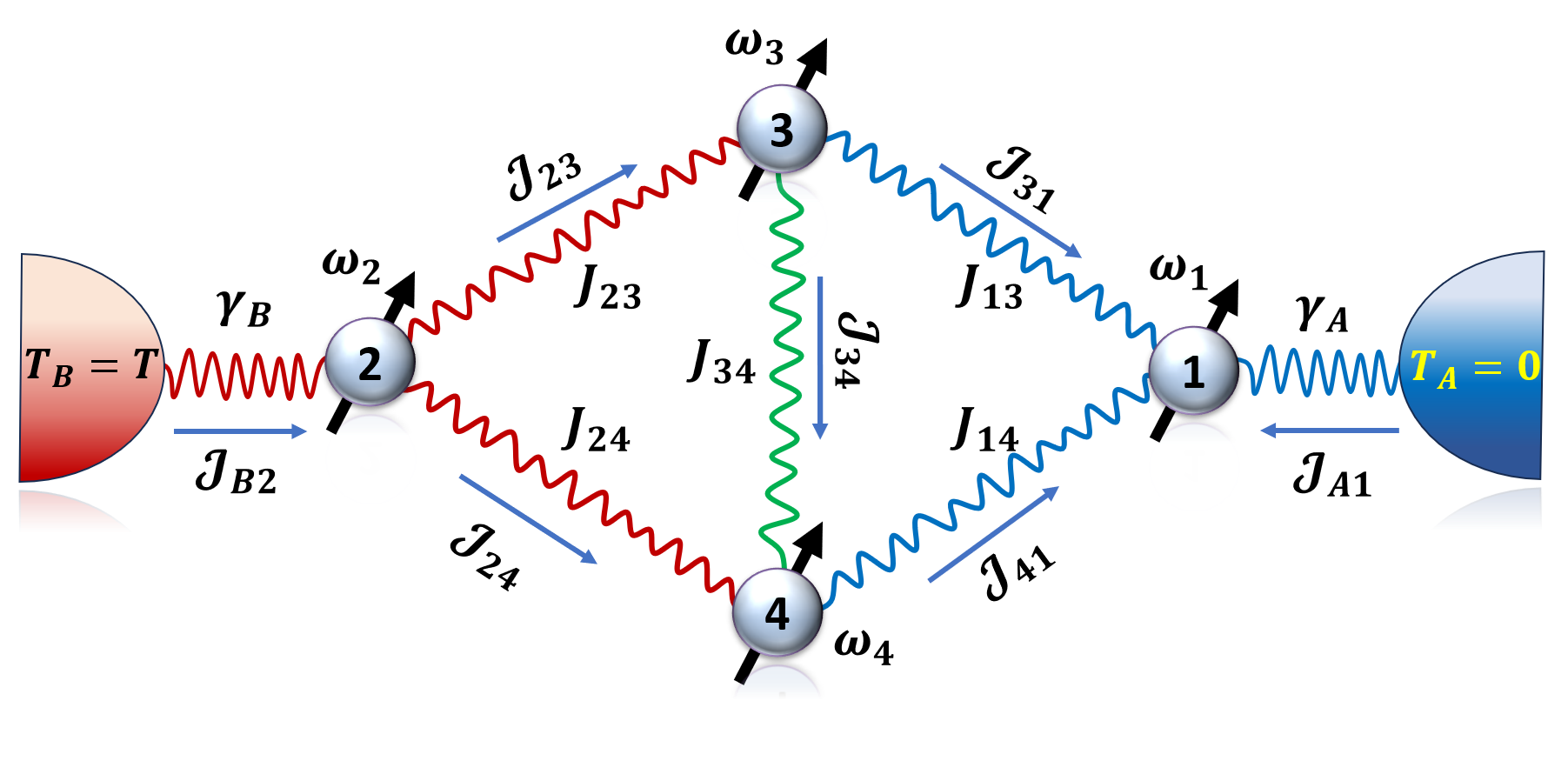}
    \caption{A schematic diagram of the standard quantum thermal Wheatstone bridge.}
    \label{fig_standard_Wheatstone_bridge}
\end{figure}
\begin{figure}
    \centering
    \includegraphics[width = 1\columnwidth]{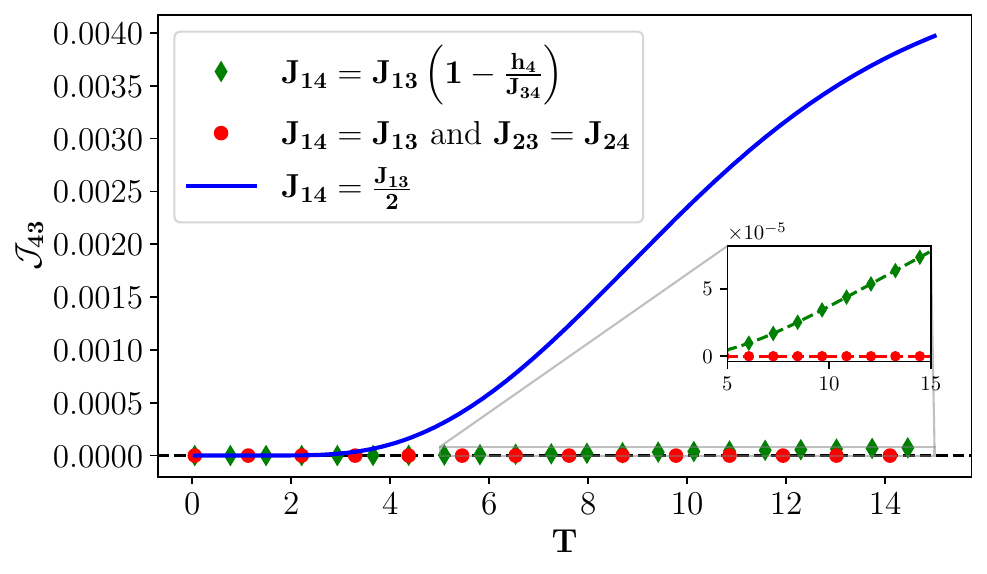}
    \caption{Variation of quantum heat current $\mathcal{J}_{43}$ between qubits 3 and 4 in the case of the quantum Wheatstone bridge model in the steady state. The green diamonds show the condition derived in~\cite{QWB}, and the red circles show the balance condition discussed in this paper. The parameters are kept the same as in~\cite{QWB} for comparison: $\omega_1 = \omega+2h_1, \omega_2 =\omega_3 = \omega, \omega_4 = \omega + 2h_4$; $\omega = 20, J_{34} = 20J, h_1 = 20J, h_4 = 0.5J, \gamma_A = J, \gamma_B = 10J, J_{23} = J_{24} = J, J_{13} = 2J$; and $J = 0.1$, where $h_i$'s are an offset in the magnetic field acting on the $i$-th spin.}
    \label{fig_QvsT_wheatstone_bridge_2}
\end{figure}

The Hamiltonian for the quantum Wheatstone bridge is given by 
$H_S^W = \frac{1}{2}\sum_{i = 1}^4\omega_i\sigma^z_i + \sum_{l, k = 1, l<k}^4J_{lk}\left(\sigma^x_l\sigma^x_k + \sigma^y_l \sigma^y_k\right)$,
where $J_{12} = 0$ as the qubits 1 and 2 are non-interacting.
A schematic diagram of the quantum thermal Wheatstone bridge is shown in Fig.~\ref{fig_standard_Wheatstone_bridge}. This is similar to the setup of a Wheatstone bridge of an electrical circuit, where an unknown resistance can be determined using a specific balance condition when the current between nodes 3 and 4 is zero. Qubits 1 and 2 are impacted by the baths $A$ at zero temperature and $B$ at temperature $T,$ respectively. The evolution of the system is dictated by the quantum master equation of the form given in Eq.~\eqref{qtr_master_eq} by replacing $H_S$ with $H_S^W$, where the steady-state $\rho^{SS}_W$ can be obtained using $\mathcal{L}\left(\rho^{SS}_W\right)=0$.
The balance condition for the Wheatstone bridge is that the quantum heat current $\mathcal{J}_{34}$ between qubits 3 and 4 should be zero. This happens when couplings follow the relation: $J_{13} = J_{14}$ and $ J_{23} = J_{24}$. Note that $J_{13}$ need not be equal to $J_{23}$.
The phenomenon of the quantum Wheatstone Bridge is shown in Fig.~\ref {fig_QvsT_wheatstone_bridge_2}. It can be observed that for the condition $J_{13} = J_{14}$, the red markers, depicting the current between qubits 3 and 4, are always zero. Further, green markers show the balance condition derived in~\cite{QWB}, which is consistent at low temperatures; however, at high temperatures, the green markers shift slightly above the zero line, deviating from the balance condition. This analysis shows that the balance condition of the Wheatstone bridge obtained in this work is more robust, and it holds at higher temperatures as well.

\section{Details on the results of the quantum thermal adder circuit}
Here, we derive Eq.~\eqref{eq_thermal_adder_result2} of the main manuscript. Consider the following relation between the currents in a quantum thermal adder circuit, derived in Eq.~\eqref{eq_thermal_adder_result1},
\begin{align}
\mathcal{J}_{A\alpha} = -\omega_\alpha\left(\frac{\mathcal{J}_{I1}}{\omega_1} + \frac{\mathcal{J}_{II2}}{\omega_2}+...+\frac{\mathcal{J}_{Nn}}{\omega_n}\right).
\label{eq_append_thermal_adder1}
\end{align}
Now, using the discussion below Eq.~\eqref{eq_relation_between_current_and_voltage} in the manuscript, we can write a relation between the effective temperatures $T_\alpha, T_1, T_2, \dots, T_n$ of the qubits $\alpha, 1, 2, \dots, n$ connected to baths $A, I, II, \dots, N$ (with temperatures $T_A, T_I, T_{II}, \dots, T_N$) and the currents $\mathcal{J}_{A\alpha}, \mathcal{J}_{I1}, \mathcal{J}_{II2}, \dots, \mathcal{J}_{Nn}$ [between (bath, qubit) $= (A, \alpha), (I, 1), (II, 2), \dots, (N, n)$, respectively] as
\begin{align}
    \mathcal{J}_{jk} = \gamma_j\omega_k\left[-1 + \coth\left(\frac{\omega_k}{2 T_j}\right)\tanh\left(\frac{\omega_k}{2 T_k}\right)\right],
    \label{eq_append_thermal_adder2}
\end{align}
for $j = A, I, II, ...N$ and $k = \alpha, 1, 2, ...n$. Upon substituting Eq.~\eqref{eq_append_thermal_adder2} in~\eqref{eq_append_thermal_adder1}, keeping in mind $T_A = 0$, we get 
\begin{align}
    \frac{-\gamma_A\omega_\alpha}{1 + e^{\omega_\alpha/T_\alpha} } &= -\frac{\omega_\alpha\gamma_I}{2}\left[-1 + \coth\left(\frac{\omega_1}{2T_I}\right)\tanh\left(\frac{\omega_1}{2T_1}\right)\right] \nonumber \\
    &-\frac{\omega_\alpha\gamma_{II}}{2}\left[-1 + \coth\left(\frac{\omega_2}{2T_{II}}\right)\tanh\left(\frac{\omega_2}{2T_2}\right)\right] + \dots\,.
\end{align}
We now take the following assumptions on parameters: $\gamma_A = \gamma_I =...=\gamma_N$, $\omega_1=\omega_2 =...=\omega_n = \omega$, $J_{1\alpha} = J_{2\alpha}=...=J_{n\alpha}$ and $T_I = T_{II}=...=T_{N} = T$. This leads to
\begin{align}
    \frac{1}{1 + e^{\omega_\alpha/T_\alpha}} &= \frac{1}{2}\left[-1 + \coth\left(\frac{\omega}{2T}\right)\tanh\left(\frac{\omega}{2T_1}\right)\right] \nonumber \\
    &+ \frac{1}{2}\left[-1 + \coth\left(\frac{\omega}{2T}\right)\tanh\left(\frac{\omega}{2T_2}\right)\right] + \dots\,,
\end{align}
which can be simplified to 
\begin{align}
    \tanh\left(\frac{\omega}{2T}\right)\left(N + \frac{2}{1 + e^{\omega_\alpha/T_\alpha}}\right) = \tanh\left(\frac{\omega}{2T_1}\right) + \tanh\left(\frac{\omega}{2T_2}\right) + \dots\,.
\end{align}
Further, in the above parameter regime, the effective temperatures $T_1, T_2, \dots, T_n$ of the qubits $1, 2, \dots, n$ are the same, say $T_x$. Using this, we can rewrite the above equation as 
\begin{align}\label{eq_append_thermal_adder_eq9}
    \tanh\left(\frac{\omega}{2T}\right)\left(N + \frac{2}{1 + e^{\omega_\alpha/T_\alpha}}\right) = N \tanh\left(\frac{\omega}{2T_x}\right),
\end{align}
which is Eq.~\eqref{eq_thermal_adder_result2} in the paper. 

We now examine the low- and high-temperature approximations of the above equation. In the low-temperature regime where $\frac{\omega}{2T}\gg 1$, we have $\tanh\left(\frac{\omega}{2T}\right)\approx 1$. Since $T_x\le T$, this simplifies to
\begin{align}
    N + \frac{2}{1 + e^{\omega_\alpha/T_\alpha}} = N,
\end{align}
implying
\begin{align}
    \frac{2}{1 + e^{\omega_\alpha/T_\alpha}} \rightarrow 0.
\end{align}
In the high-temperature regime where $\frac{\omega}{2T}\ll 1$, we have $\tanh\left(\frac{\omega}{2T}\right)\approx \frac{\omega}{2T}$. Substituting this into Eq.~\eqref{eq_append_thermal_adder_eq9}, we obtain 
\begin{align}
    \frac{\omega}{2T}\left(N + \frac{2}{1 + e^{\omega_\alpha/T_\alpha}}\right) = N\frac{\omega}{2T_x}.
\end{align}
Approximating the exponential term as $e^{\omega_\alpha/T_\alpha} \approx 1 + \frac{\omega_\alpha}{T_\alpha}$, we get 
\begin{align}
    \frac{2}{1 + e^{\omega_\alpha/T_\alpha}} = \frac{2}{2 + \frac{\omega_\alpha}{T_\alpha}} = \frac{1}{1 + \frac{\omega_\alpha}{2T_\alpha}}.
\end{align}
Using the series expansion $\frac{1}{1 + x/2} = 1 - \frac{x}{2}$ for $|x| \ll 1$, we obtain
\begin{align}
    T_x = \frac{2N~T~T_\alpha}{2 T_\alpha(N+1) - \omega_\alpha}.
\end{align}
Now, the thermal potential difference between the bath $\mathcal{B}$ and qubits $V_{\mathcal{B}x} = T- T_x$ is given by 
\begin{align}
   V_{\mathcal{B}x} = T- T_x = T \left(\frac{1 - \frac{\omega_\alpha}{2T_\alpha}}{N + 1 - \frac{\omega_\alpha}{2T_\alpha}}\right).
\end{align}
Since $\frac{\omega_\alpha}{T_\alpha}\ll 1$ in the high-temperature regime and for small $\omega_\alpha$, this simplifies to 
\begin{align}
    V_{\mathcal{B}x} \approx \frac{T}{N+1}.
\end{align}

\end{document}